\documentclass[a4paper,aps,prd,twocolumn,showpacs,eqsecnum,amsmath,amssymb,nofootinbib,superscriptaddress]{revtex4-1}
\usepackage{bm}
\usepackage{graphics}
\usepackage{graphicx}
\usepackage{epsfig}
\usepackage{booktabs,multirow}
\usepackage{hhline}
\usepackage{slashed}
\usepackage{rccol}
\usepackage{mathrsfs}
\usepackage{float}

\usepackage{xcolor}\colorlet{linkequation}{blue}
\usepackage[colorlinks=true, 
            linkcolor=red,   
            filecolor=red,   
            citecolor=green, 
            urlcolor=blue,
            hyperfootnotes=true]{hyperref}

\newcommand*{\SavedEqref}{}
\let\SavedEqref\eqref
\renewcommand*{\eqref}[1]{%
  \begingroup
    \hypersetup{
     linkcolor=linkequation,
      linkbordercolor=linkequation,
    }%
    \SavedEqref{#1}%
  \endgroup
}

\begin{document}
\newcommand{\newc}{\newcommand}
\renewcommand{\thefootnote}{\fnsymbol{footnote}}

\newc{\neutralino}{\widetilde\chi^0}
\newc{\chargino}{\widetilde\chi^{\pm}}
\newc{\squarkk}{\widetilde{q}_k}
\newc{\squarkl}{\widetilde{q}_l}
\newc{\tanb}{\tan\beta}
\newc{\gev}{\mbox{~GeV}}
\newc{\tev}{\mbox{~TeV}}

\title{Search for neutralino pair production at the CERN LHC}
\author{M.~Demirci}
\email{mehmetdemirci@ktu.edu.tr}
\affiliation{Department of Physics, Karadeniz Technical University, 61080 Trabzon, Turkey}%
\author{A.~I.~Ahmadov}
\email{ahmadovazar@yahoo.com}
\affiliation{Department of Theoretical Physics, Baku State University, Z. Khalilov Street 23,\\ AZ-1148, Baku, Azerbaijan}%
\date{March 17, 2014}
\begin{abstract}

We provide the next-to-leading order (NLO) predictions for the neutralino pair production
via quark-antiquark annihilation and gluon-gluon fusion at the CERN Large Hadron Collider,
focusing on the lightest neutralino which is likely to be the lightest supersymmetric particle.
The dependence of total LO, NLO cross sections, and \textit{K} factor on the center-of-mass energy, the $M_2$-$\mu$ mass plane,
the squark mass, and the factorization and renormalization scales
is comprehensively analyzed for three different scenarios in the minimal supersymmetric standard model
and the constrained minimal supersymmetric standard model.
We find that the LO cross section is considerably increased by the NLO correction,
and the \textit{K} factor value is clearly related to the Higgsino/gaugino mass parameters,
the squark mass, and the factorization and renormalization scales.
\end{abstract}


\maketitle

\section{\bf Introduction}
Weak scale supersymmetry (SUSY) (see, e.g., \cite{Haber,Nilles,wss}) naturally involves an elegant mechanism
for stabilizing the gauge hierarchy with regard to the effects of
radiative corrections and allows unification of gauge
couplings. Under the conservation of \textit{R}-parity\footnote{\textit{R}-parity, which is a discrete and multiplicative symmetry, is defined by $P_R=(-1)^{2S+3(B-L)}$
where $B$, $L$ and $S$ denote the baryon number, lepton number, and spin of the particle, respectively \cite{Fayet,Kazakov}. Thus, this quantity is equal to
$P_R = +1$ for the particles of the Standard Model (including the Higgs bosons) and $P_R = -1$ for their
superpartners.}, it also provides a candidate
for the dark matter (DM) postulated to explain astrophysical observations \cite{DM}. In \textit{R}-parity-conserving SUSY models, the supersymmetric particles (sparticles) can only be produced in pairs, and the lightest sparticle (LSP) is absolutely stable.
Among all the supersymmetric models, the minimal supersymmetric standard model (MSSM) is one of the most well-motivated and well-studied extensions of the standard model. The MSSM predicts many such new particles as sleptons, squarks, gluinos,
the light/heavy neutral scalar (CP-even) Higgs bosons $h^0$/$H^0$, a pseudoscalar (CP-odd)
Higgs boson $A^0$, a couple of charged Higgs bosons $H^\pm$, four neutralinos $\neutralino_{i}$
and two charginos $\chargino_j$. The neutralinos and charginos are the mass eigenstates formed from the superposition of the neutral or charged superpartners of the electroweak gauge bosons and Higgs doublets (the so-called gauginos and Higgsinos, respectively).
The lightest neutralino $\neutralino_{1}$ is usually supposed to be a weakly interacting massive particle
which is consistent with the observations of the DM candidate (see, e.g., \cite{Jungman,Griest}) in the form of the LSP for a number of SUSY breaking models. Therefore, it has to emerge as the final particle of the decay chain of each sparticle. That is why, a detailed analysis of the lightest neutralino is quite important to the phenomenological and theoretical viewpoints of SUSY.

The experimental searches of the supersymmetric particles turn out to be one of the primary tasks of the experimental
program at hadron colliders, especially at the Large Hadron Collider (LHC), after the recent discovery
of the Higgs-like boson with a mass about $126\gev$ \cite{Higgs_ATLAS,Higgs_CMS} is consistent with
the MSSM-predicted range for mass of the lightest scalar Higgs $h^0$.
Moreover, the discovery (or exclusion) of weak-scale SUSY is reckoned among the
highest physics priorities for the future LHC, including its high luminosity upgrade. Up to now, a great number of SUSY searches at
the LHC have only exhibited null results related to discovery of any supersymmetric particles. In spite of the negative results, SUSY retains strong arguments in its favour as mentioned before. These searches which chiefly focus on the production of the colored superpartners such as squarks and gluinos have been performed by the ATLAS and CMS collaborations. Consequently, new stronger limits on the masses of the first two generations squarks and gluinos have been produced depending on details of the assumed parameters.
These limits for a data set of an integrated luminosity
of around 20 fb$^{-1}$ having been collected in 8 TeV pp collisions at the LHC are given in the following.
According to recent ATLAS results \cite{sqgl_ATLAS1,sqgl_ATLAS2},
a gluino mass is excluded up to $1.1-1.3\tev$ in a mSUGRA/constrained MSSM (CMSSM) scenario at high values of the universal scalar mass parameter $m_0$ and in the gluino simplified models. The first two generations squark masses up to $700-780\gev$ are also excluded in the squark simplified models.
In addition, gluinos and squarks of equal mass are excluded for masses below
1.7 TeV in mSUGRA/CMSSM models. According to recent CMS results \cite{sqgl_CMS1}, the squark masses below 750 GeV and gluino masses of up to 1.1 TeV are excluded in the case where the squarks (gluinos) decay to one jet
(two jets) and the LSP. Owing to these stronger limits on the masses of the squarks and gluinos, the attention in the experimental researches of the supersymmetric particles starts to turn towards the electroweak production of the sleptons, neutralinos,
and charginos.

On the other hand, naturalness suggests that masses of charginos, neutralinos and third generation sparticles
should be a few hundreds of GeV range \cite{Chan}. There are also searches for superpartners of gauge and Higgs bosons, but they depend significantly on their assumed compositions and decay modes~\cite{Han}.
The bound on the lightest neutralino mass is
given by $m_{\neutralino_1} \gtrsim 46 \gev$ at 95\% CL, derived from the lower bound on chargino mass in
the MSSM at the Large Electron Positron \cite{Abdallah}. In the framework of
the CMSSM including both sfermion and gaugino mass unification, this bound reaches to well above 100 GeV from the powerful constraints set by the recent LHC data \cite{PDG}.

Note that a detailed study of the production of the lightest neutralino
$\neutralino_{1}$ and the next-to-lightest neutralino
$\neutralino_{2}$ can provide significant information about the
SUSY-breaking mechanism and the nature of the dark matter.  Moreover, the pair production of neutralinos/charginos begins to come into question as a "discovery channel" of supersymmetry. Presently one of the gold-plated SUSY discovery channels is the production of $\widetilde{\chi}_{1}^{\pm}\widetilde{\chi}_{2}^{0}$ pairs decaying into trilepton final states. But, in case  of higgsino LSP scenarios, for example appear in context of natural SUSY models, those trilepton searches loose efficiency and should be replaced by novel same-sign dilepton and 4-lepton searches~\cite{Baer}.

It is known that the effect of higher-order contributions to cross section usually increases with increment of colliding energy and would be more significant at very high energies. For this reason, it is important to take into account one-loop contributions for neutralino pair production.
In the present work we analyze the dependence of the neutralino pair production via the processes $pp(q\bar{q})\to
\neutralino_{i} \neutralino_{j}$ at tree and one-loop
levels, and $pp(\text{g}\text{g})\to\neutralino_{i} \neutralino_{j}$
at one-loop level on SUSY model parameters at the LHC energies, considering the allowed parameter
region in the MSSM.
There have been few papers dedicated to the investigations of these processes at one-loop level in literature
as follows. Considering next-to-leading order (NLO) SUSY-QCD corrections, the direct production channels of charginos and neutralinos at the Tevatron and LHC, $p \bar p /pp \rightarrow \widetilde{\chi_i}\widetilde{\chi_j}+ X$ have been worked in Ref.~\cite{Beenakker}. It has been inferred from Ref.~\cite{Beenakker} that the SUSY-QCD corrections are positive, increasing the mass range of corresponding particles that can be covered at these colliders by as much as percent 10.
The neutralino pair production via gluon-gluon fusion in the framework of the mSUGRA has been investigated
in Ref.~\cite{Yi}, and this loop-mediated process has been concluded to be competitive
with the quark-antiquark annihilation process. However, our results in present work have not exhibited this case depending on details of the SUSY model parameters. The neutralino pair production via
quark-antiquark annihilation within MSSM for three different scenarios has been worked in Ref.~\cite{Ahmadov}. The pair production of neutralinos via quark-antiquark
annihilation including the leading-log one loop radiative corrections and via gluon-gluon fusion at one-loop level (this process was computed with a numerical code) have been studied in Ref.~\cite{Gounaris}. The NLO SUSY-QCD corrections to
the production of a pair of the lightest neutralinos in association
with one jet in the framework of the phenomenological MSSM (p19MSSM) have been computed in Ref.~\cite{Cullen}. Finally, recently in our previous paper (see Ref.~\cite{Demirci2}) we have also analyzed the leading and subleading electroweak (EW) corrections to the neutralino pair production at proton-proton collision, and we have found that the EW corrections supply sizeable contributions, in particular, for the process $pp\to \neutralino_{2} \neutralino_{2}$.

Unlike the above-mentioned works, within the present work the most outstanding feature of our
approach is the mechanism in selecting the input parameters. We recover the corresponding
Lagrangian parameters as direct analytical expressions of appropriate physical masses
without any restrictions on them in the MSSM. As a matter of fact, we mainly focus on
the algebraically nontrivial inversion in order to obtain Higgsino and gaugino mass parameters.
If we need to explicitly specify, we can say that using $\tan\beta$ and masses of charginos as input
parameters, then we get the other ones being Higgsino/gaugino mass parameters,
neutralino masses and mixing matrix.

The remainder of the present work proceeds in the following order: In
Section \ref{sec:qqbarninj}, the analytical results of the relevant amplitudes and cross sections
are given for partonic process $q\bar{q}\to \widetilde\chi_{i}^{0}\widetilde\chi_{j}^{0}$. In Section
\ref{sec:oneloop}, we give briefly information about one-loop contributions to neutralino pair production
via quark-antiquark annihilation (in Subsection \ref{sec:qqbarone}) and gluon-gluon
fusion (in Subsection \ref{sec:ggone}). In Section \ref{sec:input}, we present definitions corresponding
to our method and input parameters which are used in numerical calculations. In Section \ref{sec:results},
we give numerical results and discuss the corresponding SUSY parameters dependences of the cross section in detail for each scenario. Finally, the results appearing in Section~\ref{sec:results} are summarized in Section~\ref{sec:conclusion}.

\section{The Leading-Order Calculation for The Neutralino Pair Production}\label{sec:qqbarninj}
In this section, after introducing the necessary couplings and Lagrangians in the MSSM,
we serve up analytical results of amplitudes and cross section for the partonic process $q\overline
q\rightarrow\neutralino_{i}\neutralino_{j}$ at leading order (LO).
The clean environment of proton-proton collision, together with the well-defined energy of the initial state,
make this collision ideal for precision measurements of neutralinos properties. The associated production of
neutralino pair via quark-antiquark collision at hadron colliders could be denoted by
\begin{equation}\label{eq:qqninj}
q(p_1)\overline q(p_2)\rightarrow\neutralino_{i}(k_1)\neutralino_{j}(k_2),
\end{equation}
where the labels in parentheses indicate the four momenta of the relevant particles. The cross section for subprocess~\eqref{eq:qqninj} is parameterized in terms of the following Mandelstam variables,
\begin{equation}
\hat s=(p_1+p_2)^2,~\hat t=(p_1-k_1)^2,~\hat
u=(p_1-k_2)^2.
\end{equation}
Introducing by ($\theta,p$) scattering angle and momentum in the
center-of-mass system of the final states neutralinos, for corresponding center-of-mass
energy and momentums we have,
\begin{equation} \label{eq:cm}
\begin{split}
 p&=\frac{1}{2 \sqrt{\hat s}}\sqrt{(\hat s-m_{\neutralino_{i}}^2-m_{\neutralino_{j}}^2)^2-4m_{\neutralino_{i}}^2
m_{\neutralino_{j}}^2},\\
 E_1&=\frac{\hat s+m_{\neutralino_{i}}^2-m_{\neutralino_{j}}^2}{2 \sqrt{\hat s}},~E_2=\frac{\hat
s+m_{\neutralino_{j}}^2-m_{\neutralino_{i}}^2}{2 \sqrt{\hat s}},\\
 p_1&=\frac{\sqrt{\hat s}}{2}(1,0,0,1),~p_2=\frac{\sqrt{\hat
s}}{2}(1,0,0,-1),\\
 k_1&=(E_1,p \sin\theta,0,p \cos\theta),\\
k_2&=(E_2,-p \sin\theta,0,-p
\cos\theta).
\end{split}
\end{equation}
In the following part, we give the corresponding couplings of the neutralino pair production
in the MSSM. Using the standard
notation, the $Z^0$ boson-neutralino-neutralino interactions are
proportional to the following couplings:
\begin{equation} \label{eq:Oij}
\begin{split}
& O_{ij}^{''L}=\frac{1}{2}\left[ N_{i4}N_{j4}^* - N_{i3}N_{j3}^*\right],\\
& O_{ij}^{''R}=\frac{1}{2}\left[ N_{i3}^*N_{j3} - N_{i4}^*N_{j4}\right],\\
\end{split}
\end{equation}
where $O_{ij}^{''R}=-O_{ij}^{''L*}$, and $N$ denotes neutralino mixing matrix being
a $4\times4$ unitary matrix which diagonalizes
the neutralino mass matrix. Neglecting generational mixing in the squark sectors, then,
the neutralino-quark-squark interactions
are proportional to the relevant couplings,
\begin{equation} \label{eq:CNqsq}
\begin{split}
C_{\neutralino_{i} \squarkk q}^L&=\bigl[(e_q-I_q^3)s_W N_{i1}+ I_q^3 c_W N_{i2}\bigr]\delta_{kL}\\
&+\frac{c_W m_q(N_{i4}~\delta_{qu} + N_{i3}~\delta_{qd})}{2m_W(\sin\beta~\delta_{qu}+\cos\beta~\delta_{qd})}\delta_{kR},\\
C_{\neutralino_{i} \squarkk q}^R&=(-e_q s_W N_{i1}^*) \delta_{kR}\\
&+\frac{m_q c_W(N_{i4}^*~\delta_{qu}
+N_{i3}^*~\delta_{qd})}{2m_W(\sin\beta~\delta_{qu}+\cos\beta ~\delta_{qd})}\delta_{kL},\\
\end{split}
\end{equation}
and for the $Z^0$ boson-quark-quark couplings, we have
\begin{equation}
\begin{split}
&C_{Z q q}^L=2I_{q}^3(1-2s^2_{W}|e_q|),\\
&C_{Z q q}^R=-2s^2_{W}e_q,
\end{split}
\end{equation}
where $e_q$ and $I_q^3$ are the fractional electromagnetic charge and the third
component of the weak isospin of quark $q$; such that $I_{q_L}^3=\pm 1/2~(I_{q_R}^3=0)$ for left-handed
(right-handed) up- and down-type quarks. The sine and cosine of the electroweak mixing angle $\theta_W$ are
denoted by $c_W \equiv \cos\theta_W=m_W/m_Z$ and $s_W\equiv\sin\theta_W=\sqrt{1-c_W^2}$. In the above couplings,
furthermore, $q$ refers to up- and down-type quarks, while the label $k$ refers to left- and right-handed
for squark. Finally, $\delta_{kl}$ appearing in Eq.~\eqref{eq:CNqsq} is the kronecker delta function which is equal to 1 if the labels $k,l$ are the same, and 0 otherwise; for instance $\delta_{qu}=1$ for up-type quark ($q\equiv u$) and $\delta_{kL}=0$ for
right-handed squark ($k\equiv R$). We use it to display the neutralino couplings to
both an up-type quark/squark and a down-type quark/squark in the same relation. The couplings of the neutralino to $Z^0$ boson and
(s)quark are considerably dependent on the corresponding elements of the neutralino mixing matrix
$N_{ij}$ ($i,j=1,...,4$) as seen from the above couplings. Considering neutralino mass eigenstate basis,
the neutralino interactions to corresponding particles in question are obtained from the following Lagrangians \cite{Haber},
\begin{equation}  \label{eq:LqsqN}
\mathcal{L}_{\neutralino_{i} \squarkk q}= -\frac{\sqrt2 g}{c_W}
\overline{\neutralino_{i}} q \left[C_{\neutralino_i \squarkk q}^{L*}P_{L}
+ C_{\neutralino_i \squarkk q}^{R*}P_{R}\right]\squarkk,
\end{equation}
\begin{equation} \label{eq:LZNN}
\mathcal{L}_{Z^{0}\neutralino_{i}\neutralino_{j}}=\frac{g}{c_W} Z_\mu \overline{\neutralino_{i}}
\gamma^{\mu}\left[O_{ij}^{''L} P_{L}+O_{ij}^{''R}P_{R}\right]\neutralino_{j},
\end{equation}
\begin{equation} \label{eq:LZqq}
\mathcal{L}_{Z^{0}q\bar{q}}=-\frac{g}{2c_{W}} \bar{q}\gamma^{\mu}
\left[ C_{Z q q}^L P_{L}+C_{Z q q}^R P_{R}\right]q Z_{\mu},
\end{equation}
where $q$, $\squarkk$ and $\neutralino_{i}$ are four-component spinor
fields of the quark, squark and neutralino, respectively; $P_{R,L}=\frac{1}{2}(1\pm\gamma^5)$
are the chiral projectors; and $g=e/s_W$ is the $SU(2)$ gauge coupling. Note that the Higgsino and
gaugino components of the neutralino in the $Z\neutralino_{i}\neutralino_{j}$ and
$q \squarkk\neutralino_{i}$ coupling are controlled by the neutralino mixing matrix as shown in the above Lagrangians.

\begin{figure*}[!htp]
    \begin{center}
\includegraphics[scale=1.43]{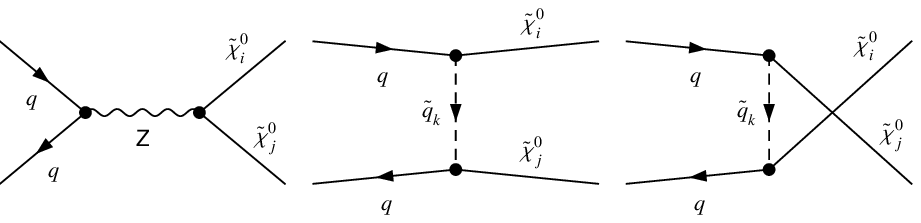}
     \end{center}
\caption{Feynman diagrams of the partonic process $q\bar{q}
\to\neutralino_{i}\neutralino_{j}$ at leading level.}\label{fig:fig1}
\end{figure*}

The Feynman diagrams of the partonic process $q\bar{q}
\to\neutralino_{i}\neutralino_{j}$ at leading level are displayed in Fig.~\ref{fig:fig1}.
We neglect the contributions from the Feynman diagrams including the couplings $H^0/G^0/A^0-q-q$ seeing that the strength of Yukawa coupling is proportional to the fermion mass and masses of the first two generations quarks are relatively small and could be ignored. Nevertheless, we will take into account these couplings and contributions of this vertex for bottom quark in a further work. Consequently, the subprocess for neutralino pair production contains an $s$-channel contribution through exchanging the $Z^0$ boson,
$t$- and $u$-channel contributions via exchanging of the squarks as shown in Fig.~\ref{fig:fig1}.
The leading-level contributions to the amplitude emerging from the three channels are given by
\begin{align}  \label{eq:Ts}
T_{\hat s}=&\frac{-g^2 D_{Z}(\hat
s)}{2c_W^2}\biggl[\overline{u}_{i}(k_1) \gamma_{\mu}(O^{''L}_{ij}P_L +
O^{''R}_{ij}P_R){v}_{j} (k_2)\biggr]\nonumber \\
&\cdot\biggl[\overline{v}(p_2)\gamma^{\mu}\left(C_{Z q q}^L P_L+C_{Z q q}^R P_R\right)u(p_1)\biggr],
\end{align}
\begin{align}  \label{eq:Tt}
T_{\hat t}=&\sum_{k} \frac{2g^2}{(\hat t-m_{\squarkk}^2)c_W^2}
\biggl[\overline{u}_i(k_1)
(C_{\neutralino_i \squarkk q}^{L}P_L +C_{\neutralino_i \squarkk q}^{R}P_R) u(p_1)\biggr]\nonumber \\
&\cdot\biggl[\overline{v}(p_2)
(C_{\neutralino_j \squarkk q}^{R*} P_{L}+C_{\neutralino_j \squarkk q}^{L*}P_{R})
v_j(k_2)\biggr],
\end{align}
\begin{align}  \label{eq:Tu}
T_{\hat u}=&\sum_{l}\frac{-2g^2}{(\hat u-m_{\squarkl}^2)c_W^2}
\biggl[\overline{u}_j(k_2)
(C_{\neutralino_j {\squarkl} q}^{L} P_{L}+C_{\neutralino_j {\squarkl} q}^{R}P_{R})
u(p_1)\biggr]\nonumber \\
&\cdot\biggl[\overline{v}(p_2)
(C_{\neutralino_i \squarkl q}^{R*} P_{L}+C_{\neutralino_i \squarkl q}^{L*}P_{R})
v_i(k_1)\biggr],
\end{align}
where the labels $k,l$ represent the summation over the exchanged left/right-handed components of squarks
in the same flavor, and the labels $i,j$ represent the type of the neutralinos in the final state.
From the above amplitudes along with couplings~\eqref{eq:Oij} and~\eqref{eq:CNqsq}, explicitly we note that
purely Higgsino production dominates in the contribution coming from the $s$-channel diagram,
whereas the $t$- and $u$-channel contributions are dominated by purely gaugino production. After averaging over colors and spins of incoming particles, the parton-level differential cross section in the analytic form is given
by the following formula,
\begin{align}
\frac{d\hat \sigma (q\overline
q\rightarrow\neutralino_{i}\neutralino_{j})}{d\hat t}&=\frac{1}{16\pi \hat
s^2}\frac{1}{12}\left(\frac{1}{2}\right)^{\delta_{ij}} (M_{\hat s \hat s}+M_{\hat t \hat t}+M_{\hat u \hat u}\nonumber \\
&-2M_{\hat s \hat t}+2M_{\hat s \hat u}-2M_{\hat t \hat u}),
\end{align}
where the factors $\frac{1}{12}$ is arising from spin and color
averaging over the initial state and $(\frac{1}{2})^{\delta_{ij}}$ denotes the final identical particle factor.
Using standard trace techniques, the squared amplitudes explicitly take the following form,

\begin{align} \label{eq:Mss_udbarNC}
M_{\hat s \hat s} =& \frac{g^4|D_{Z}(\hat s)|^2}{c_W^4}
[(C_{Z q q}^L)^2+(C_{Z q q}^R)^2]
\biggl\{O^{''L}_{ij}O^{''L*}_{ij}\nonumber \\
&\times[(m_{\neutralino_{i}}^2 - \hat u)(m_{\neutralino_{j}}^2 -\hat u)
+(m_{\neutralino_{i}}^2 - \hat t)(m_{\neutralino_{j}}^2 -\hat t)]\nonumber \\
&-[|O^{''L}_{ij}|^2+|O^{''R}_{ij}|^2]m_{\neutralino_{i}}m_{\neutralino_{j}}\hat s\biggr\},
\end{align}
\begin{align}
M_{\hat t \hat t} =& \sum_{k,l}\frac{4g^4}{(\hat t - m_{\squarkk}^2)(\hat t-m_{\squarkl}^2)c_W^4}
\biggl\{(m_{\neutralino_{i}}^2- \hat t)(m_{\neutralino_{j}}^2 - \hat t)\nonumber \\
&\times[C_{\neutralino_i \squarkk q}^{L}C_{\neutralino_i \squarkl q}^{L*}+ C_{\neutralino_i \squarkk q}^{R}C_{\neutralino_i \squarkl q}^{R*}]\nonumber \\
&\times[C_{\neutralino_j \squarkk q}^{L}C_{\neutralino_j \squarkl q}^{L*}
   +C_{\neutralino_j \squarkk q}^{R}C_{\neutralino_j \squarkl q}^{R*}]\biggr\},
\end{align}
\begin{align}
M_{\hat u \hat u} =& \sum_{k,l}\frac{4g^4}{(\hat u - m_{\squarkk}^2)(\hat
u-m_{\squarkl}^2)c_W^4} \biggl\{(m_{\neutralino_{i}}^2- \hat u)(m_{\neutralino_{j}}^2 - \hat u)\nonumber \\
&\times[C_{\neutralino_i \squarkk q}^{L*}C_{\neutralino_i \squarkl q}^{L}
+C_{\neutralino_i \squarkk q}^{R*}C_{\neutralino_i \squarkl q}^{R}]\nonumber \\
&\times[C_{\neutralino_j \squarkk q}^{L*}C_{\neutralino_j \squarkl q}^{L}
  +C_{\neutralino_j \squarkk q}^{R*}C_{\neutralino_j \squarkl q}^{R}]\biggr\},
\end{align}
\begin{align}
M_{\hat t \hat u}=&\sum_{k,l}\frac{4g^4}{(\hat t -
{m}_{\squarkk}^2)(\hat u-m_{\squarkl}^2)c_W^4}
\biggl\{\frac{1}{2}[C_{\neutralino_i \squarkk q}^{L*}
C_{\neutralino_j \squarkl q}^{L} C_{\neutralino_i \squarkl q}^{R*}
C_{\neutralino_j \squarkk q}^{R}\nonumber \\
&+C_{\neutralino_i \squarkl q}^{L*}
C_{\neutralino_j \squarkk q}^{L} C_{\neutralino_i \squarkk q}^{R*}
C_{\neutralino_j \squarkl q}^{R}]
[(m_{\neutralino_{i}}^2-\hat u)(m_{\neutralino_{j}}^2-\hat u)\nonumber \\
&+(m_{\neutralino_{i}}^2-\hat t)(m_{\neutralino_{j}}^2-\hat t)-\hat s(\hat s
-m_{\neutralino_{i}}^2-m_{\neutralino_{j}}^2)]\nonumber \\
&+m_{\neutralino_{i}} m_{\neutralino_{j}}\hat
s [C_{\neutralino_j \squarkl q}^{L*}
C_{\neutralino_i \squarkk q}^{L} C_{\neutralino_j \squarkk q}^{L*}
C_{\neutralino_i \squarkl q}^{L}\nonumber \\
&+C_{\neutralino_j \squarkl q}^{R*}
C_{\neutralino_i \squarkk q}^{R} C_{\neutralino_j \squarkk q}^{R*}
C_{\neutralino_i \squarkl q}^{R}]\biggr\},
\end{align}
\begin{align}
M_{\hat s \hat u}=&\sum_{k} \frac{2g^4 (Re[D_{Z}(\hat s)])}{(\hat
u-m_{\squarkk}^2)c_W^4}\biggl\{[C_{Z q q}^L
O^{''L*}_{ij} C_{\neutralino_i \squarkk q}^{L*} C_{\neutralino_j \squarkk q}^{L}\nonumber \\
&-C_{Z q q}^R O^{''L}_{ij} C_{\neutralino_i \squarkk q}^{R*} C_{\neutralino_j \squarkk q}^{R}
](m_{\neutralino_{i}}^2-\hat u)(m_{\neutralino_{j}}^2 -\hat u)\nonumber \\
&+ m_{\neutralino_{i}} m_{\neutralino_{j}} \hat s [C_{Z q q}^R O^{''L*}_{ij} C_{\neutralino_i \squarkk q}^{R*} C_{\neutralino_j \squarkk q}^{R}\nonumber \\
&- C_{Z q q}^L O^{''L}_{ij} C_{\neutralino_i \squarkk q}^{L*} C_{\neutralino_j \squarkk q}^{L} ]\biggr\},
\end{align}
\begin{align}
M_{\hat s \hat t}=&\sum_{k} \frac{2g^4 (Re[D_{Z}(\hat s)])}{(\hat t-m_{\squarkk}^2)c_W^4}
\biggl\{[C_{Z q q}^R O^{''L*}_{ij} C_{\neutralino_j \squarkk q}^{R*} C_{\neutralino_i \squarkk q}^{R}\nonumber \\
& - C_{Z q q}^L O^{''L}_{ij} C_{\neutralino_j \squarkk q}^{L*} C_{\neutralino_i \squarkk q}^{L} ](m_{\neutralino_{i}}^2-\hat t)(m_{\neutralino_{j}}^2 -\hat t) \nonumber \\
&+ m_{\neutralino_{i}} m_{\neutralino_{j}} \hat s [C_{Z q q}^L O^{''L*}_{ij} C_{\neutralino_j \squarkk q}^{L*} C_{\neutralino_i \squarkk q}^{L}\nonumber \\
&- C_{Z q q}^R O^{''L}_{ij} C_{\neutralino_j \squarkk q}^{R*} C_{\neutralino_i \squarkk q}^{R}]
\biggr\},
\end{align}
where $D_{Z}(\hat s)$ is propagator of the $Z^0$ boson.

For obtaining the total cross section of the subprocess we use
the following formula:
\begin{equation} \label{eq:sigma}
\hat \sigma(\hat s)=\int_{\hat{t}^-}^{\hat{t}^+}d\hat t ~\frac{d\hat
\sigma}{d\hat t},
\end{equation}
where the upper and lower bounds of integral are defined as $\hat{t}^\pm=1/2\bigl[(m_i^2+m_j^2-\hat s)
\pm\sqrt{(\hat s-m_i^2-m_j^2)^2-4m_i^2 m_j^2}\bigr]$. Once the cross section for the partonic process
has been computed, the total hadronic cross sections in proton-proton collisions
in terms of the center-of-mass energy could be readily obtained using
\begin{equation} \label{eq:totalsigma}
\sigma(s)=\int_{(m_{\neutralino_{i}}+m_{\neutralino_{j}})^2/s}^{1}d\tau~
\frac{d\mathcal{L}^{AB}_{ab}}{d\tau}
~\hat{\sigma}(\text{\textit{subprocess, at}}~ \hat s=\tau s),
\end{equation}
with the parton luminosity
\begin{align} \label{eq:pluminosity}
\frac{d\mathcal{L}^{AB}_{ab}}{d\tau}&=\int_\tau^1
\frac{dx_1}{x_1}\frac{1}{1+\delta_{ab}}\biggl[G_{a/A}(x_1,\mu_F)G_{b/B}(\frac{\tau}{x_1},\mu_F)\nonumber \\
&+G_{b/A}(x_1,\mu_F)G_{a/B}(\frac{\tau}{x_1},\mu_F)\biggr],
\end{align}
where the universal parton distribution functions (PDFs) for
the partons $a,b$, constituents of hadrons $A,B$ are denoted by $G_{a/A}$ and $G_{b/B}$, depending on the
longitudinal momentum fractions of the two partons $x_1,x_2$
($\tau=x_1 x_2$) at a factorization scale $\mu_F$. During our calculations, the factorization scale is chosen
as the average mass of the produced particles, namely, $\mu_F=(m_{\neutralino_{i}}+m_{\neutralino_{j}})/2$.

\section{One-Loop Contributions to the neutralino pair production}\label{sec:oneloop}
At the one-loop level production of neutralino pair is proceeded via
quark-antiquark annihilation and gluon-gluon fusion in the hadron colliders. Feynman diagrams for the one-loop
contributions to the process $pp\rightarrow\neutralino_{i}\neutralino_{j}$ can be divided into three
kind diagrams as follows: The box diagrams, the self energy corrections diagrams,
and triangle diagrams.
Any one-loop amplitude could be given as a linear sum of triangle, box, bubble, and
tadpole one-loop integrals.

In the numerical calculations of high-energy processes observed at
the current and future accelerators such as LHC and ILC, for precise
theoretical predictions of cross sections one needs to include higher-order
corrections. In the common case it is explained
in the following: First of all, the lowest-order approximation in
perturbative calculations of high energy physics is not
sufficiently accurate to be compared to the experimental data.
Thus, it is important to consider the contributions from higher-order terms as
well. For including these corrections in the Standard model or
beyond, it is indispensable to handle the evaluation of loop
integrals.

We briefly describe the general properties of the box, triangle and self energy corrections diagrams in the following part.
The general form the triangle diagram in four dimensions is
proportional to the antisymmetric tensor
$\varepsilon_{\mu\nu\rho\sigma}$. Such tensor could not be continued to general
dimensions, because it has exactly four indices. Therefore, such diagram is excluded from the general
proof and has to be treated separately via a different
regularization scheme, e.g. the Pauli-Villars method. It
must be verified that  all higher-order diagrams including
the $\varepsilon_{\mu\nu\rho\sigma}$ tensor may be
renormalized without demolishing gauge invariance. One of the main
conditions for the proof of renormalizability, in general, is that this
scheme should be gauge invariant and the Slavnov-Taylor
identities can be established.

In our case self-energy diagrams consist of the quark, squark,
and boson self-energy corrections. These contributions have different
properties. It should be noted that the self-energy of the fermions is not
physically observable, and therefore it does not make sense even if
it has the logarithmic divergence. The basic problem should appear when there is
a logarithmic divergence in the evaluation of the
physical observable. The most important example is the vertex
correction due to the photon or gluon propagation. If it has a
logarithmic divergence, then it should be renormalized into the wave
function.

We have performed numerical calculations in the 't Hooft-Feynman gauge where the gluon polarization
sum is given by $\sum_\lambda \epsilon_\mu^*(k,\lambda)\epsilon_\nu(k,\lambda)=-g_{\mu\nu}$.
We have considered the constrained differential renormalization (CDR) \cite{CDR} with a view to regularize the ultraviolet (UV) divergences. At the one-loop level, the CDR has been presented to be equivalent to
the regularization by dimensional reduction \cite{DR,DR2}, which is a modified version of dimensional
regularization. Hence, a supersymmetry-preserving regularization scheme is supplied by the
implementation given in Ref. \cite{DR3}. For a treatment of the appearing infrared (IR) and collinear singularities we use mass regularization, such as IR singularities are treated by a small gluon mass, and the masses of the light quarks are kept in collinearly singular integrals.

We do not give the analytical results for the one-loop level since these are too long to
be included here. Now we give kinematic expressions and the Feynman diagrams
for the neutralino pair production in the next subsections, considering each partonic process separately.

\subsection{The partonic process $\mathbf{q\bar{q}\rightarrow\neutralino_{i}\neutralino_{j}}$ in the one-loop level}\label{sec:qqbarone}
The Feynman diagrams contributing to the subprocess $q\bar{q}\rightarrow
\neutralino_{i}\neutralino_{j}$ in the one-loop level are depicted from Fig.~\ref{fig:fig3} to \ref{fig:fig2}. The virtual corrections to this process include the following generic structure of one-loop Feynman diagrams: Self-energy, three-point vertex and box corrections as
shown in Figs.~\ref{fig:fig3},~\ref{fig:fig4} and~\ref{fig:fig2}, respectively. In these figures the label $S^0$ represents all neutral Higgs bosons $h^0, H^0, A^0, G^0$, and
the label $\widetilde{f}^w_m~({f}_m)$ refers to scalar fermions (fermions)
$\widetilde{e}^w_m, \widetilde{u}^w_m, \widetilde{d}^w_m~(e_m,\nu_m,u_m,d_m)$. The subscript $m$ and superscripts $w,x,y$ refer to the generation of (s)quark
and the squark mass eigenstates, respectively.
\begin{widetext}
\begin{figure*}[!ht]
    \begin{center}
\includegraphics[width=\textwidth,height=20 cm]{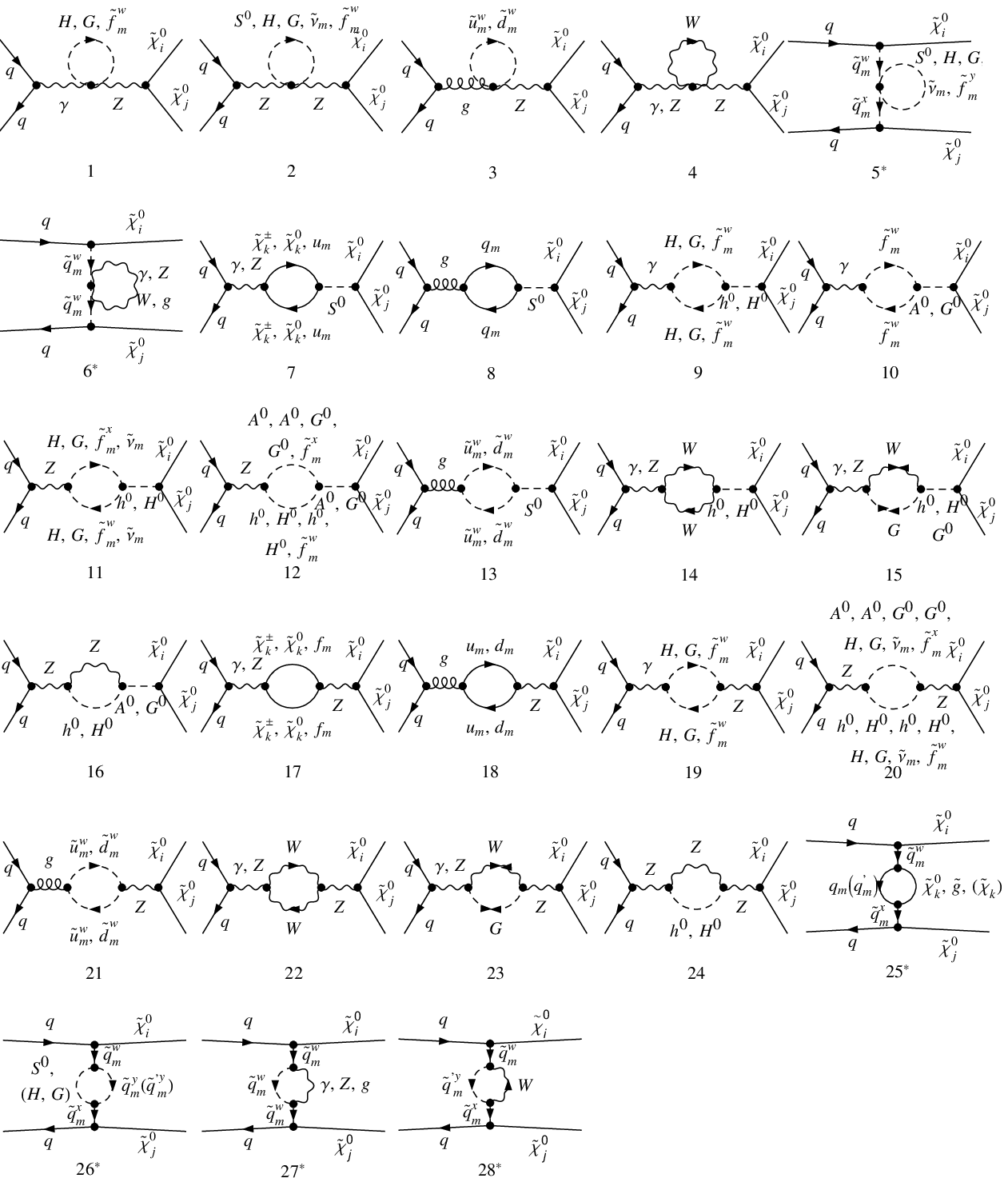}
     \end{center}
\caption{Feynman diagrams for self-energy corrections to neutralino pair production
via $q\bar{q}\rightarrow\widetilde\chi_{i}^{0}\widetilde\chi_{j}^{0}$
to one-loop level. Here, the diagrams with exchanging the final state neutralinos in the $t$-channel diagrams are
not explicitly shown. The star on the numbers under some diagrams refers to the $t$-channel diagrams.}\label{fig:fig3}
\end{figure*}
\begin{figure*}[!ht]
    \begin{center}
\includegraphics[width=\textwidth]{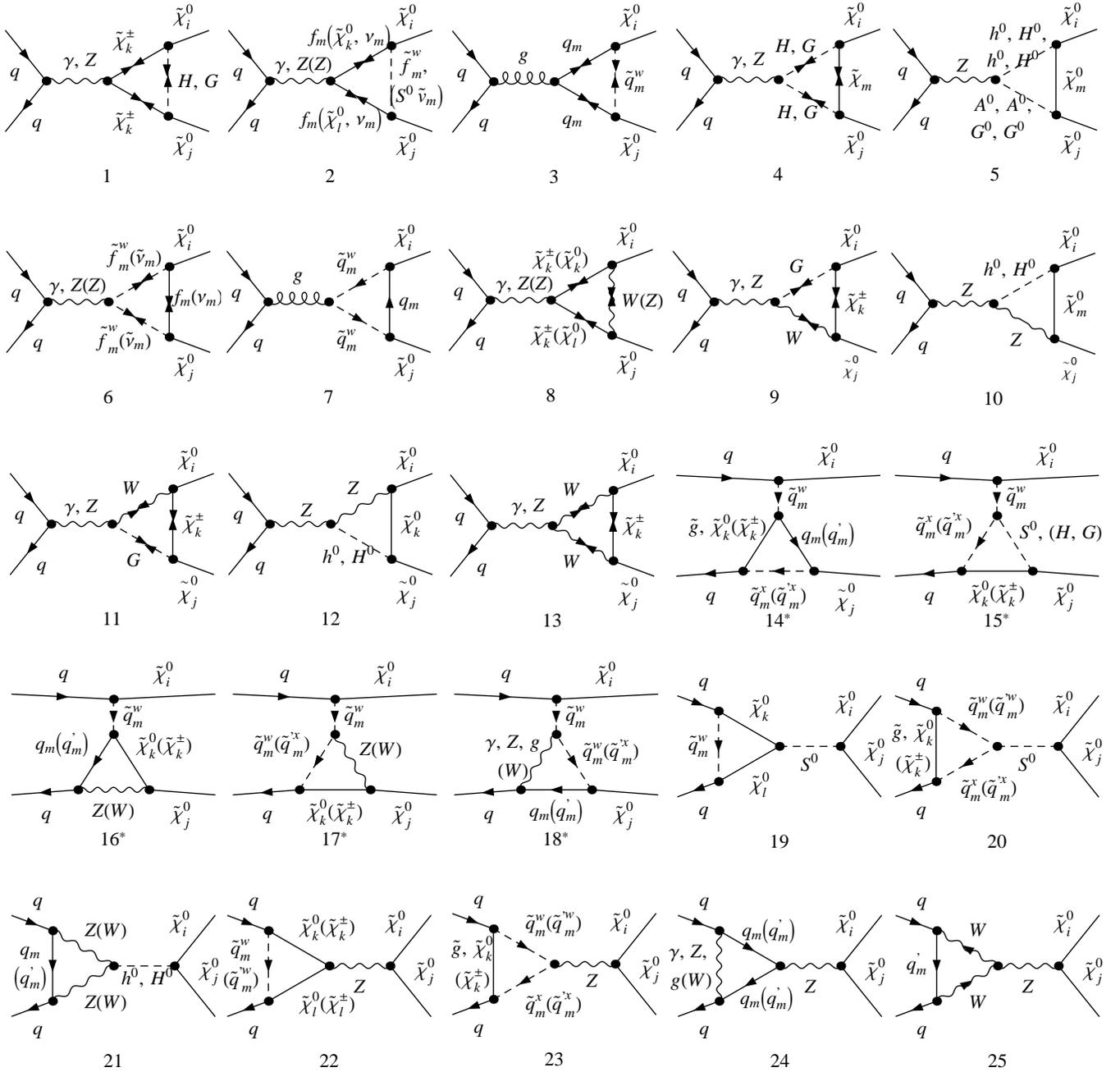}
     \end{center}
\caption{Feynman diagrams for vertex corrections to neutralino pair production
via $q\bar{q}\rightarrow\widetilde\chi_{i}^{0}\widetilde\chi_{j}^{0}$
to one-loop level. Also, this subprocess contains diagrams which have corrections in the upper vertex
including the same triangle corrections in the diagrams from 14 to 18. Here, the diagrams with exchanging the final state neutralinos in the $t$-channel diagrams are
not explicitly shown. The star on the numbers under some diagrams refers to the $t$-channel diagrams.} \label{fig:fig4}
\end{figure*}
\begin{figure*}[!ht]
    \begin{center}
\includegraphics[width=\textwidth]{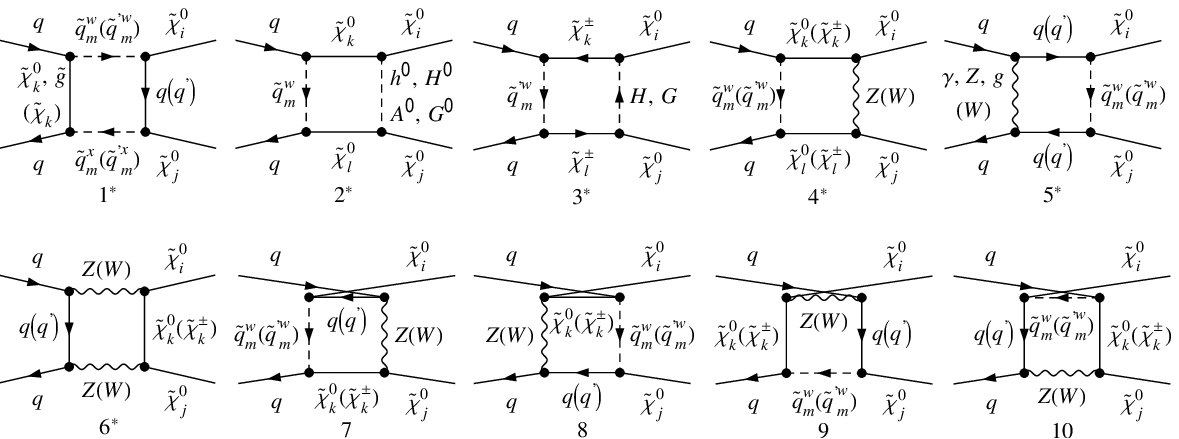}
     \end{center}
\caption{Feynman diagrams for box corrections to neutralino pair production
via $q\bar{q}\rightarrow\neutralino_{i}\neutralino_{j}$
to one-loop level. Here, the diagrams with exchanging the final state neutralinos in the $t$-channel diagrams are
not explicitly shown. The star on the numbers under some diagrams refers to the $t$-channel diagrams.}\label{fig:fig2}
\end{figure*}
\end{widetext}
We denote  the process of neutralino pair production via quark-antiquark annhilation as
\begin{equation} \label{eq:qqbarone}
q(p_1)\bar{q}(p_2)\rightarrow\widetilde\chi_{i}^{0}(k_1)\widetilde\chi_{j}^{0}(k_2),
\end{equation}
where the labels in parentheses represent the four momenta of the corresponding particles.

\begin{figure*}[!t]
    \begin{center}
\includegraphics[width=\linewidth,height=11.424 cm]{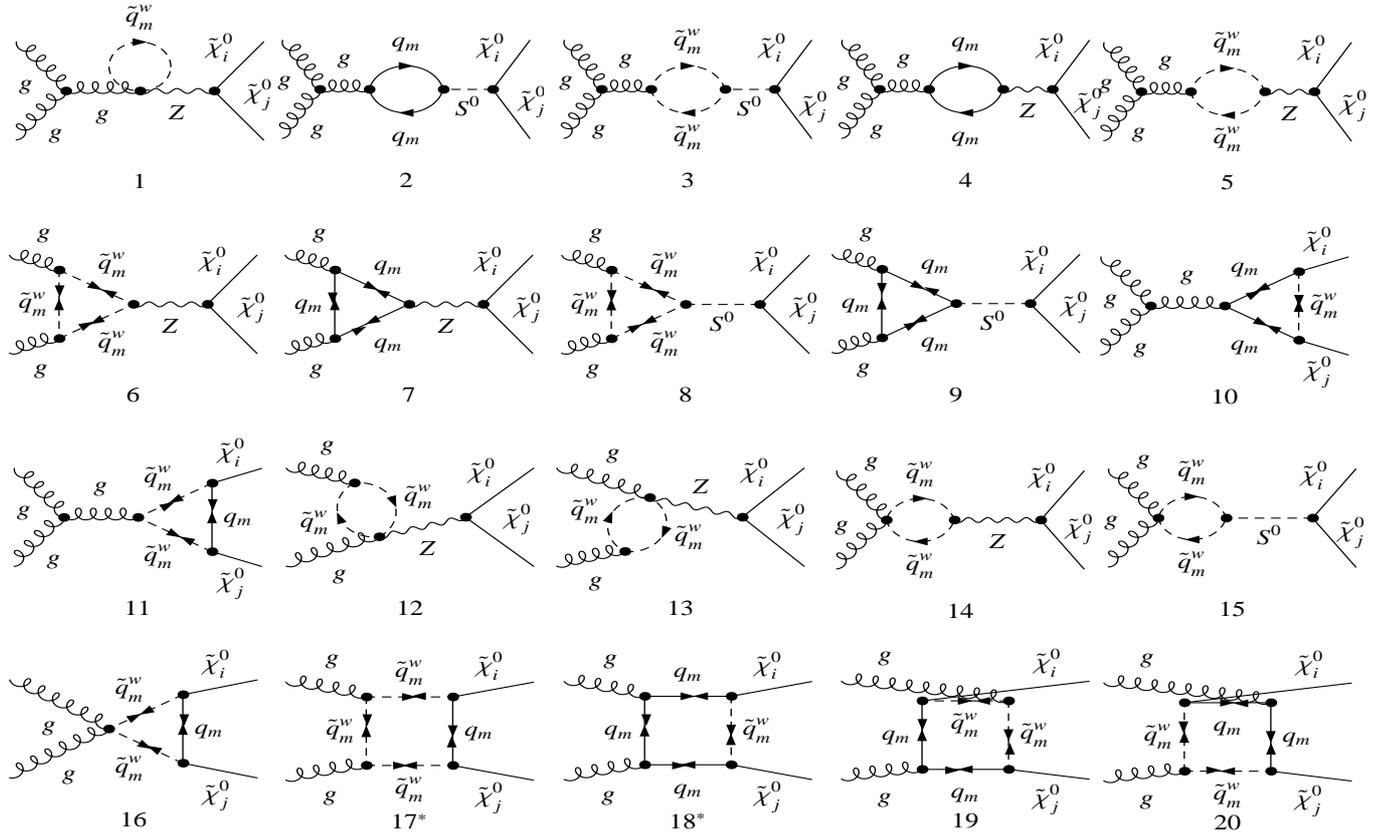}
     \end{center}
\caption{Feynman diagrams for virtual corrections to neutralino pair production
via $\text{g}\text{g}\rightarrow\widetilde\chi_{i}^{0}\widetilde\chi_{i}^{0}$
to one-loop level. Here, the diagrams with crossed final states are not explicitly shown. The subscript $m$ and superscripts $w$ refer to the generation of (s)quark
and the squark mass eigenstates, respectively.}\label{fig:fig5}
\end{figure*}

\subsection{The partonic process $\mathbf{\text{g}\text{g}\rightarrow\neutralino_{i}\neutralino_{j}}$
in the one-loop level}\label{sec:ggone}
The subprocess $\text{g}\text{g}\rightarrow \neutralino_{i}\neutralino_{j}$ in the lowest order can only be
produced by way of one-loop diagrams, namely it does not emerge at the
tree level. We represent the process of neutralino pair production via gluon-gluon fusion with
\begin{equation} \label{eq:ggnn}
\text{g}(p_3)\text{g}(p_4)\rightarrow\widetilde\chi_{i}^{0}(k_3)\widetilde\chi_{j}^{0}(k_4),
\end{equation}
where the labels in parentheses represent the four momenta of the relevant particles. The
Mandelstam variables for subprocess~\eqref{eq:ggnn} are given by
\begin{equation}
\hat s=(p_3+p_4)^2, ~\hat t=(p_3-k_3)^2,~\hat
u=(p_3-k_4)^2.
\end{equation}
For this process, there is no need to take into account the
renormalization at the one-loop level and  provided that all of the one-loop
contributions are involved in the MSSM, the UV divergence
will automatically be canceled.
The Feynman diagrams contributing to the subprocess $\text{g}\text{g}\rightarrow \neutralino_{i}\neutralino_{j}$ in the one-loop level are depicted in Fig.~\ref{fig:fig5}.
The virtual corrections to this process include the following generic structure of one-loop Feynman diagrams: Self-energy, vertex and box corrections as shown in diagrams from 1 to 5, 6 to 15 and 16 to 20 in Fig.~\ref{fig:fig5}, respectively. As seen from these diagrams, this process involves virtual quark/squark corrections. In this figure all neutral Higgs bosons $h^0, H^0, A^0, G^0$ are denoted by the label $S^0$ and the star on the numbers under some diagrams represents that these are t-channel diagrams.

\section{Parameter Space}\label{sec:input}
We now give the information about our method and input parameters used in the numerical analysis. During our numerical evaluations, we take into account the assumptions and approaches in our previous paper \cite{Demirci}
for the gaugino/Higgsino sector. The soft SUSY-breaking gaugino mass parameters $M_1$, $M_2$ and
the Higgsino mass parameter $\mu$ can be taken to be real and positive. These gaugino mass parameters are commonly
supposed to be connected by way of the relation $M_1=\frac{5}{3}M_2
\tan^2\theta_W\simeq 0.5 M_2$. The parameters $M_2$ and $\mu$ are obtained as shown in Eqs. (A13) and (A14)
in Ref. \cite{Demirci} by taking the suitable differences and sums of the chargino masses.
Consequently, there appear three different cases in the selection of the gaugino/Higgsino mass
parameters $M_2$ and $\mu$. These are the Higgsino-like, gauginolike, and mixture-case, separately.
We can refer the reader to Ref.~\cite{Demirci} for further details. We set the chargino masses as
\begin{equation} \label{eq:mc}
m_{\chargino_{1}}=168.51\gev,~m_{\chargino_{2}}= 295.01\gev
\end{equation}
for both Higgsino-like and gauginolike scenarios, and
\begin{equation} \label{eq:mc2}
m_{\chargino_{1}}= 173.66\gev,~m_{\chargino_{2}}= 289.86\gev
\end{equation}
for mixture-case scenario. Then, the parameters $\mu$ and $M_2$ related to the scenarios are calculated from
these values in~\eqref{eq:mc} and~\eqref{eq:mc2} for given $\tan\beta$. Furthermore, neutralino masses for each scenario are obtained
by inserting the values of $\mu$ and $M_2$ into Eq. (A8) in Ref. \cite{Demirci}.
Taking into account the constraint on SUSY parameters from recent experiments \cite{sqgl_ATLAS1,sqgl_ATLAS2,sqgl_CMS1},
we set the soft SUSY-breaking parameters for the entries of mass matrices
in the sfermion sector to be equal as $M_{SUSY}=1.5\tev$. We get the other SUSY parameters as follows:
\begin{equation}
\begin{split}
&\tanb=45,~m_{A^{0}} = 2.5\tev,\\
&A_t=A_b=A_{\tau}=\mu / \tanb +2M_{SUSY}, \\
&m_{\widetilde{u}_{L}}= 1499.02\gev,~m_{\widetilde{u}_{R}}= 1499.59\gev,\\
&m_{\widetilde{d}_{L}}=1500.18\gev,~m_{\widetilde{d}_{R}}= 1501.20\gev,\\
&m_{\widetilde{g}}=1500\gev,
\end{split}
\end{equation}
where $A_{t,b,\tau}$ are the trilinear couplings and $m_{A^{0}}$ is the mass of the neutral CP-odd Higgs boson.
Furthermore, we take the following input parameters
for the SM, $m_Z= 91.1876\gev$, $m_W= 80.399\gev$, $\alpha^{-1}= 137.036$, $\alpha(m_Z^2)^{-1}= 127.934$ and  $\alpha_s(m_Z^2)= 0.1184$ \cite{PDG}, and we ignore the masses of the light quarks.
The running strong coupling $\alpha_s(\mu_0^2)$ at energy scale $\mu_0=(m_{\neutralino_{1}}+m_{\neutralino_{1}})/2$
yields 0.1152, 0.1183, and 0.1165 in the Higgsino-like scenario, gauginolike scenario, and mixture-case scenario, respectively.

Additionally, we have considered the CMSSM 40.2.4 benchmark point \cite{CMSSM4022} in order to make the comparison with our scenarios.
The CMSSM \cite{CMSSM,CMSSM2,CMSSM3} contains five
input parameters, namely, the universal trilinear soft SUSY breaking
parameter $A_0$, the universal scalar mass parameter $m_{0}$, gaugino mass
parameter $m_{1/2}$, the ratio of the expectation values of the two Higgs
doublets $\tanb$ and the sign of the Higgs mixing parameter sign($\mu$).
It is believed that the universal parameters $A_0$, $m_{0}$, and $m_{1/2}$ arise via some gravity-mediated mechanism,
and these are defined at the grand unified theories scale while sign($\mu$) and $\tan\beta$ are described at the
electroweak scale.
In the CMSSM 40.2.4 benchmark point, the input parameters are given as follows: $m_0=700\gev$, $m_{1/2}=600\gev$,
$A_0=-500\gev$, $\tanb=40$, and $\mu>0$. In this case,
we obtain the corresponding SUSY particle spectrum with the help of
\verb"SoftSusy-3.3.9" package \cite{softsusy} as follows:
\begin{equation}
\begin{split}
&m_{\chargino_{1}}=480.02\gev,~m_{\chargino_{2}}=809.62\gev,\\
&m_{\widetilde{u}_{L}}=1413.98\gev,~m_{\widetilde{u}_{R}}=1374.64\gev,\\
&m_{\widetilde{d}_{L}}=1416.06\gev,~m_{\widetilde{d}_{R}}=1370.96\gev,\\
&m_{\widetilde{\text{g}}}=1384.44\gev,~m_{h^{0}} = 118.04\gev,\\
&m_{A^{0}} =m_{H^{0}}= 807.41\gev.
\end{split}
\end{equation}
\begin{table*}[ht]
\caption{The Higgsino/gaugino mass parameters, neutralino masses, and $\tanb$
for each scenario, where all mass parameters are in GeV.}\label{tab:table1}
\begin{ruledtabular}
\begin{tabular}{lR[.][.]{3}{2}R[.][.]{3}{2}R[.][.]{3}{2}R[.][.]{3}{2}R[.][.]{3}{2}R[.][.]{3}{2}R[.][.]{3}{2}R[.][.]{3}{2}}
 &\multicolumn{1}{c}{$M_{2}$}&\multicolumn{1}{c}{$\mu$}&\multicolumn{1}{c}{$M_{1}$}
 &\multicolumn{1}{c}{$\tanb$}&\multicolumn{1}{c}{$m_{\neutralino_{1}}$}& \multicolumn{1}{c}{$m_{\neutralino_{2}}$}&\multicolumn{1}{c}{$m_{\neutralino_{3}}$}&\multicolumn{1}{c}{$m_{\neutralino_{4}}$}\\
\hline
Higgsino-like&250.00&200.00&119.33&45&109.59&174.50&209.65&294.88\\
Gauginolike& 200&250.00& 95.46&45 &91.50&169.50&259.40&293.85\\
Mixture case&225.00&225.00&107.39&45 &101.42&176.13&234.52&289.37\\
CMSSM 40.2.4~~& 470.87&795.94& 254.88&40 &251.96& 479.89&800.38&808.69\\
\end{tabular}
\end{ruledtabular}
\end{table*}
Furthermore, Table~\ref{tab:table1} shows a list of the Higgsino/gaugino mass parameters, neutralino masses,
and $\tanb$ for our scenairos and the CMSSM 40.2.4 benchmark point.

\section{Numerical results and discussion}\label{sec:results}

Let us now discuss in detail the numerical predictions of the process $pp \to\neutralino_{i}\neutralino_{j}$ at the LHC
energies, taking into account the full one-loop contributions from quark-antiquark annihilation and
gluon-gluon fusion. We carry out the numerical evaluation using the
Mathematica packages \texttt{FEYNARTS} \cite{Feynarts} to obtain
the relevant amplitudes, \texttt{FORMCALC} \cite{Hahn} to
supply both the analytical results and a complete Fortran code for numerical evaluation of the squared matrix
elements, and \texttt{LOOPTOOLS} \cite{loop} to make the evaluation
of the necessary loop integrals as based on Passarino-Veltman reduction techniques \cite{PV}.
In addition, with the help of \texttt{FEYNARTS} we generate all relevant Feynman diagrams,
which are shown in Figs.~\ref{fig:fig1} through~\ref{fig:fig5}.
Higgs properties are computed by using \texttt{FEYNHIGGS} \cite{FeynHiggs}. In the numerical treatment,
we use the MSTW2008 PDFs \cite{MSTW} interfaced via the \texttt{LHAPDF} package \cite{LHAPDF} for the distribution of the gluon/quark in the proton. Moreover, we set the central renormalization and factorization scales to be equal ($\mu_0=\mu_F=\mu_R$) and fix $\mu_0$ as the average mass of the produced particles $\mu_0=(m_{\neutralino_{i}}+m_{\neutralino_{j}})/2$ in default. To have a quantitative understanding of the effects of one-loop contributions on the neutralino pair production, it is convenient to compute the \textit{K} factor, which is defined as the ratio between the total NLO and LO cross sections, namely, $\textit{K}=(\sigma_{NLO}^{q \bar q}+\sigma_{NLO}^{\text{gg}})/\sigma_{LO}$.

\begin{figure}[ht]
    \begin{center}
\includegraphics[scale=0.33]{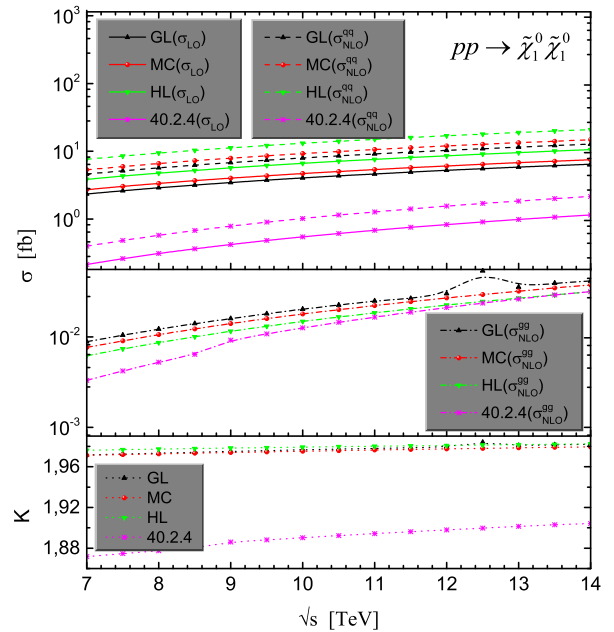}
     \end{center}
\caption{(color online). Total LO and NLO cross sections of
the process $pp\to\widetilde{\chi}_{1}^0\widetilde{\chi}_{1}^0$ versus the center-of-mass energy of $pp$ collider. The lower panel shows
the \textit{K} factor, $K=(\sigma_{NLO}^{q \bar q}+\sigma_{NLO}^{\text{gg}})/\sigma_{LO}$.} \label{fig:fig6}
\end{figure}

\begin{figure}[ttt]
    \begin{center}
\includegraphics[scale=0.33]{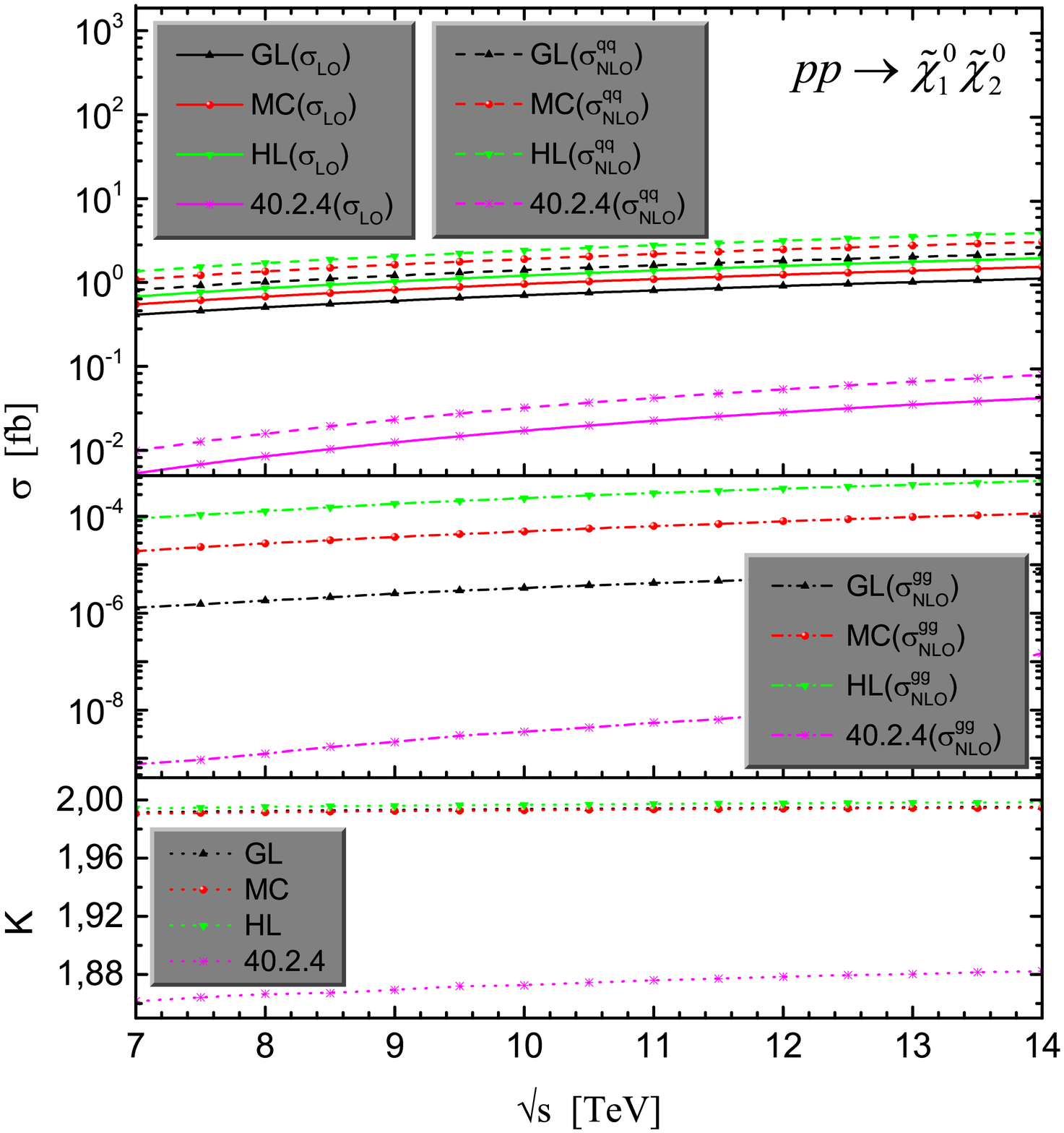}
     \end{center}
\caption{(color online). Total LO and NLO cross sections of
the process $pp\to\widetilde{\chi}_{1}^0\widetilde{\chi}_{2}^0$ versus the center-of-mass energy of $pp$ collider. The lower panel shows
the \textit{K} factor, $K=(\sigma_{NLO}^{q \bar q}+\sigma_{NLO}^{\text{gg}})/\sigma_{LO}$.}\label{fig:fig7}
\end{figure}

\begin{figure}[ttt]
    \begin{center}
\includegraphics[scale=0.33]{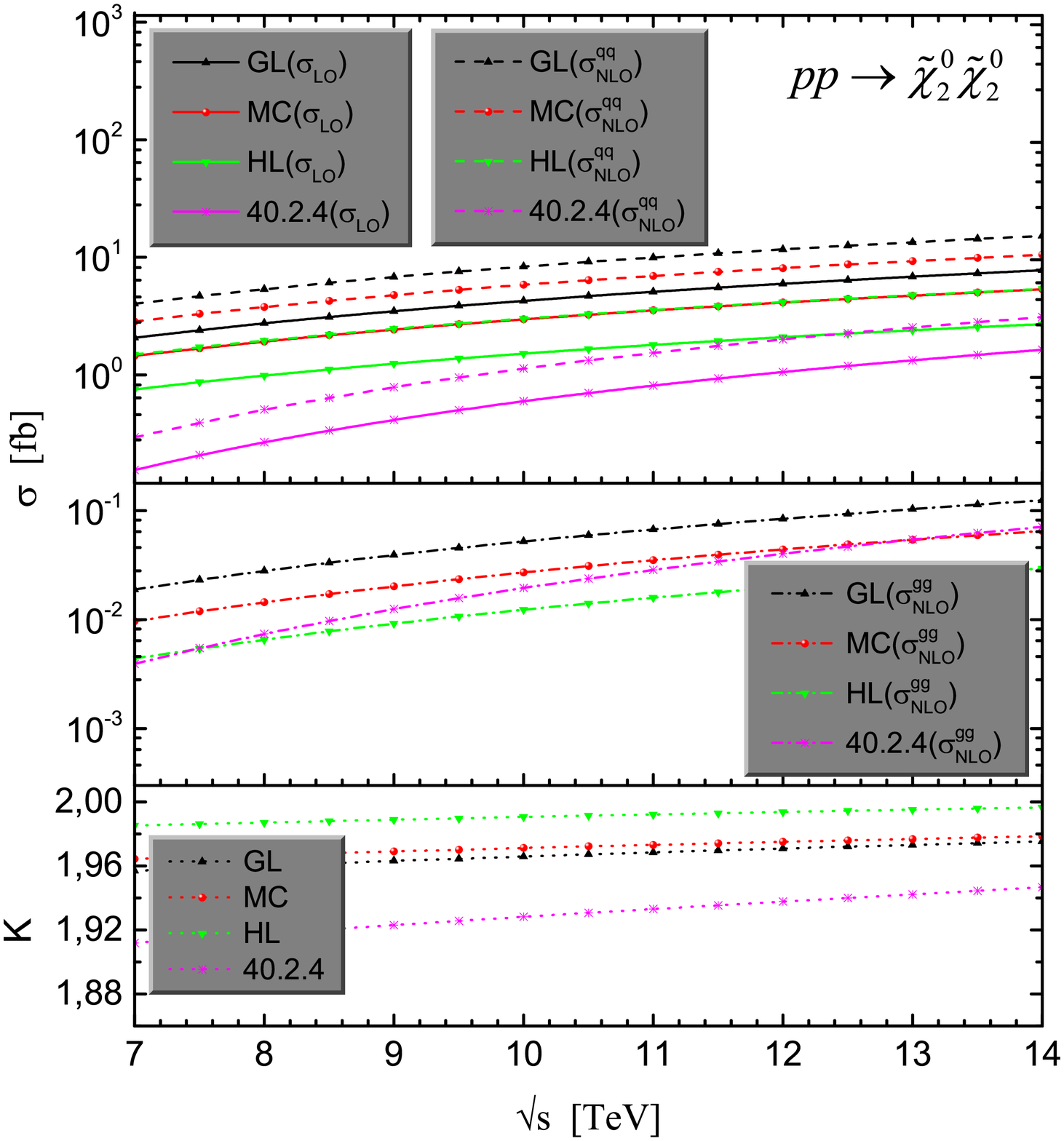}
     \end{center}
\caption{(color online). Total LO and NLO cross sections of
the process $pp\to\widetilde{\chi}_{2}^0\widetilde{\chi}_{2}^0$ versus the center-of-mass energy of $pp$ collider. The lower panel shows
the \textit{K} factor, $K=(\sigma_{NLO}^{q \bar q}+\sigma_{NLO}^{\text{gg}})/\sigma_{LO}$.} \label{fig:fig8}
\end{figure}

\begin{figure*}[!ht]
    \begin{center}
\includegraphics[scale=0.34]{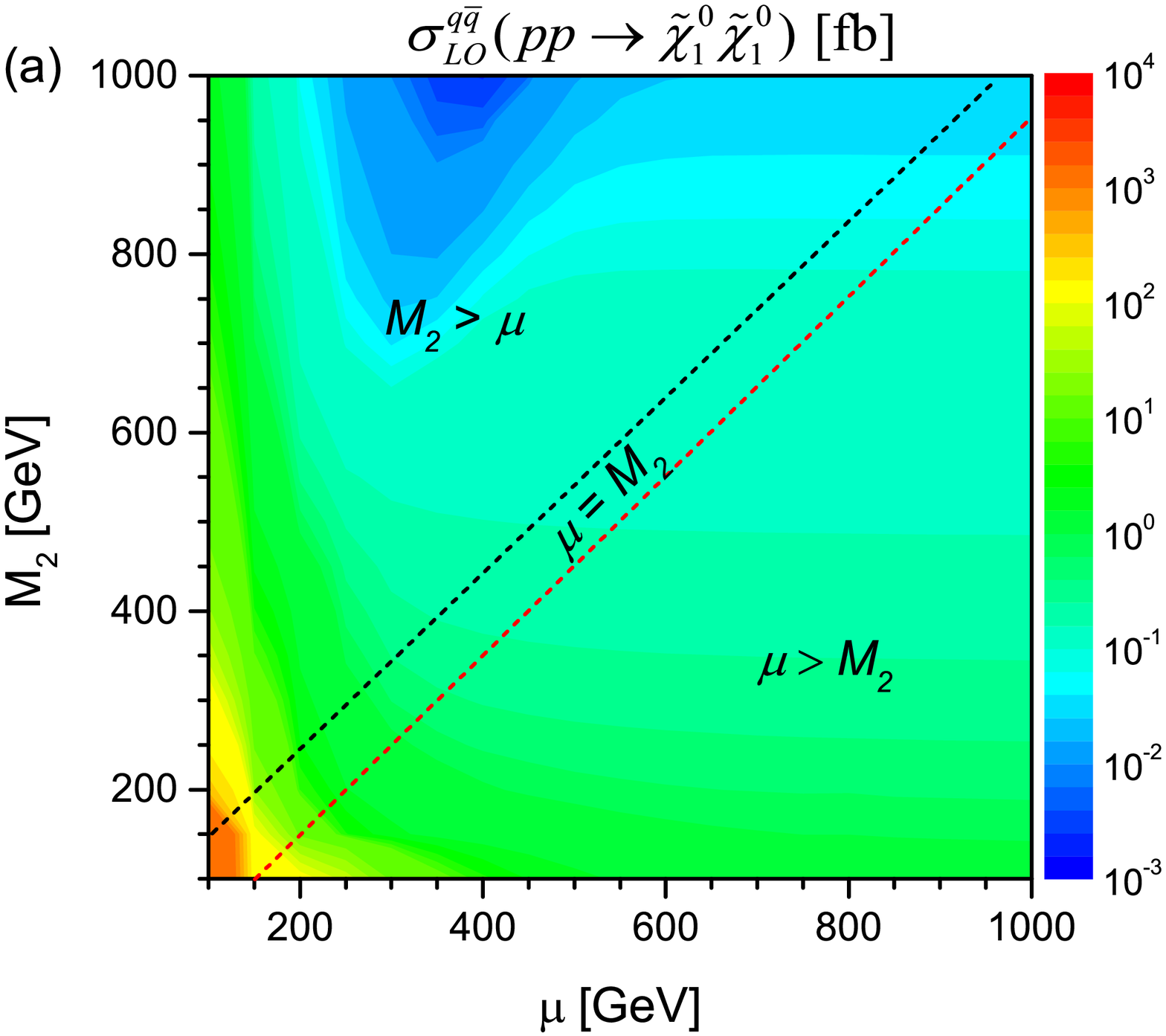}
\includegraphics[scale=0.34]{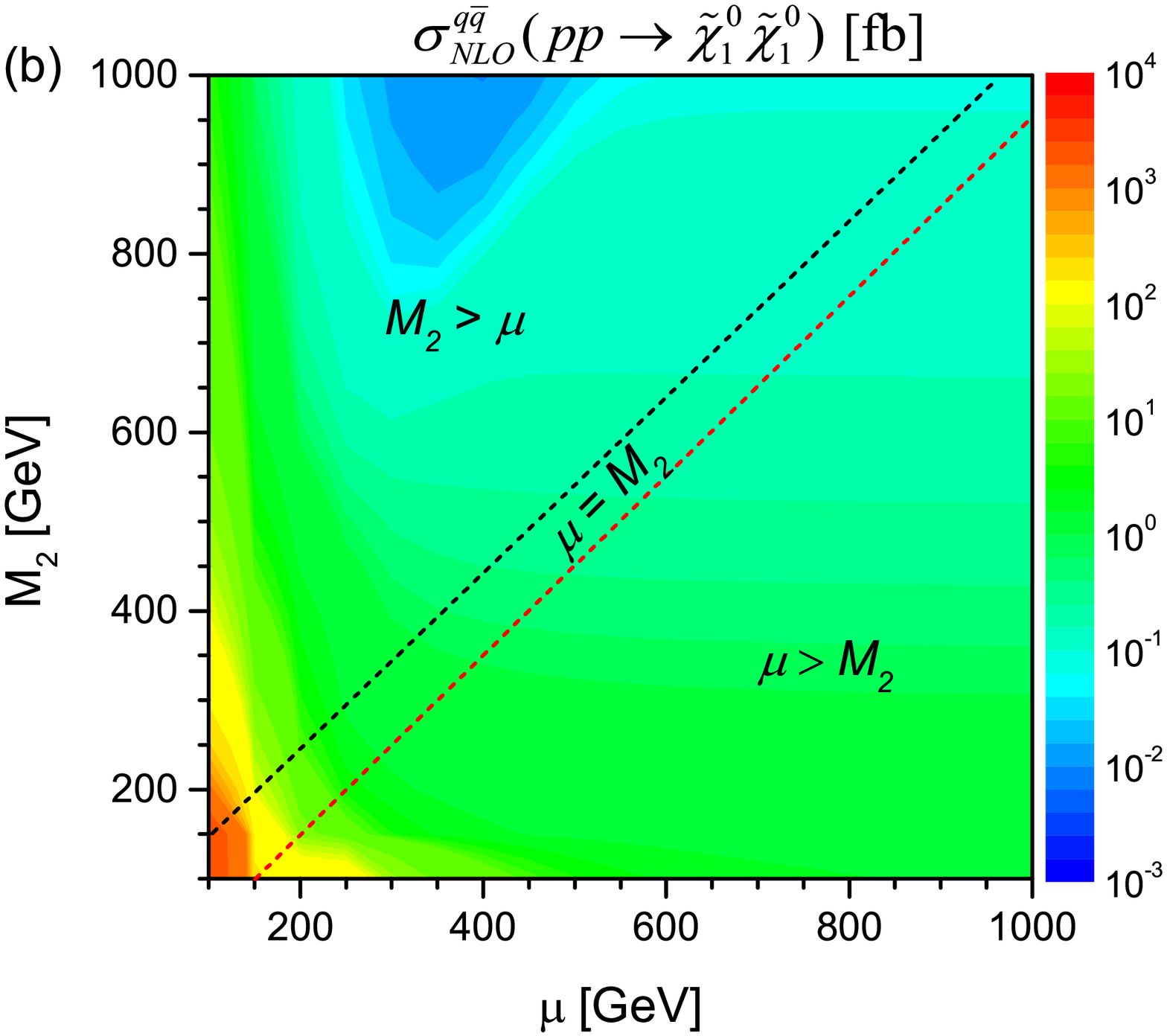}
\includegraphics[scale=0.34]{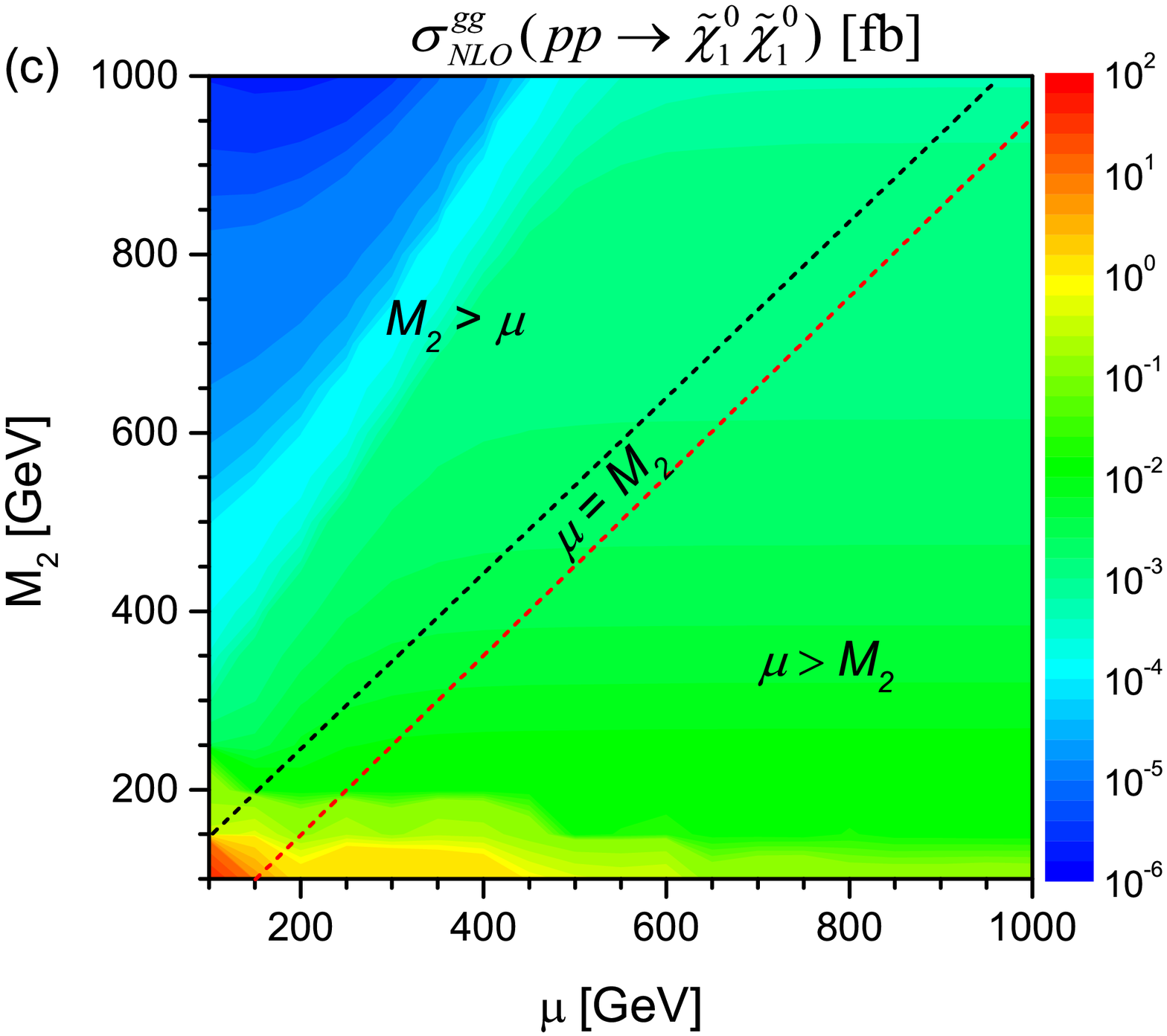}
\includegraphics[scale=0.34]{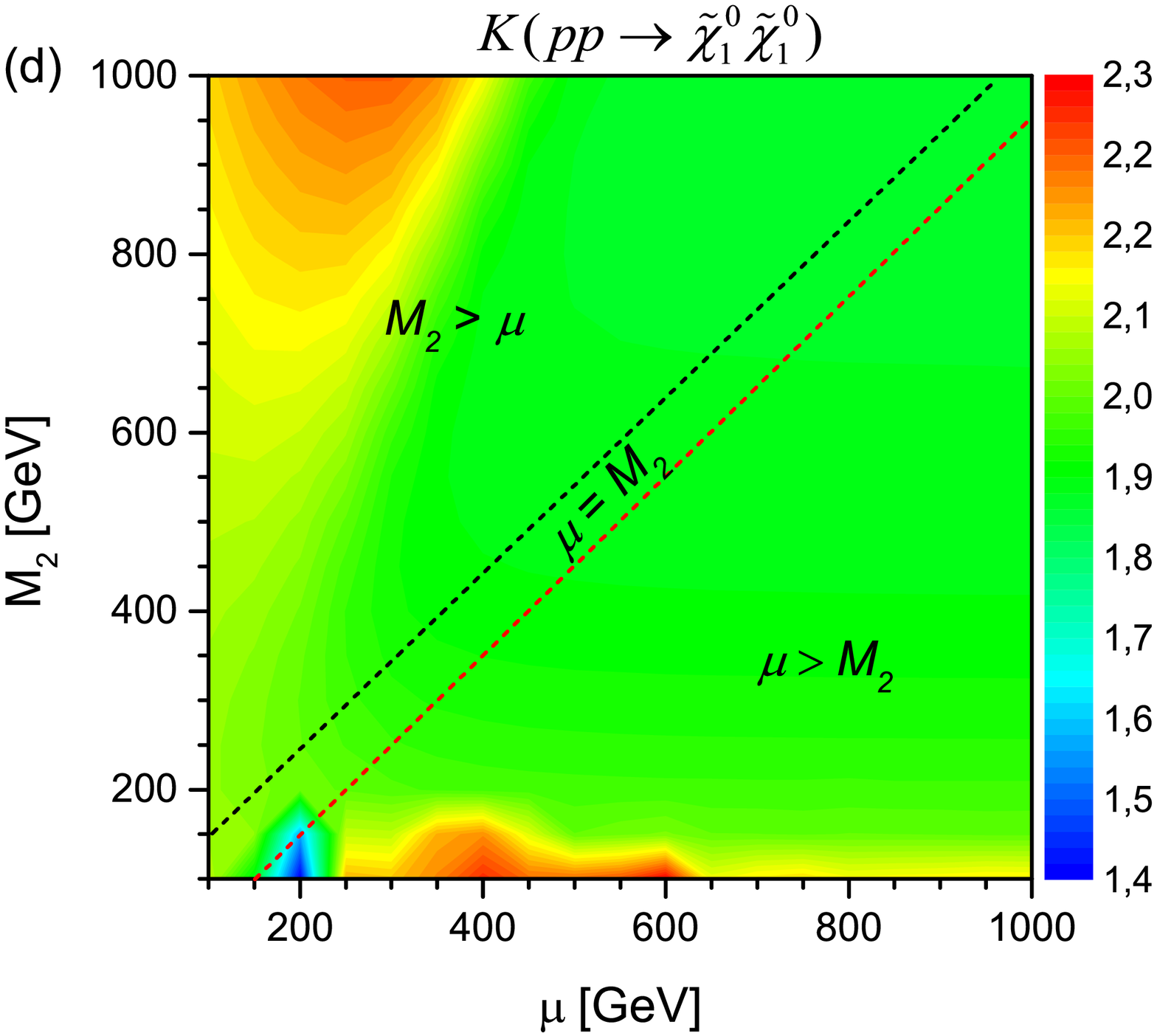}
      \end{center}
\caption{(color online). Contour plots of the total (a) LO, (b)-(c) NLO cross sections and (d) \textit{K} factor
of the process  $pp\to\widetilde{\chi}_{1}^0\widetilde{\chi}_{1}^0$ in the $M_2-\mu$ plane for $\sqrt{s}=8$ TeV,
where we take $\tanb=45$ and fix $M_1=\frac{5}{3}M_2 \tan^2\theta_W$.}
\label{fig:fig9}
\end{figure*}

\begin{figure}[hpt]
    \begin{center}
\includegraphics[scale=0.34]{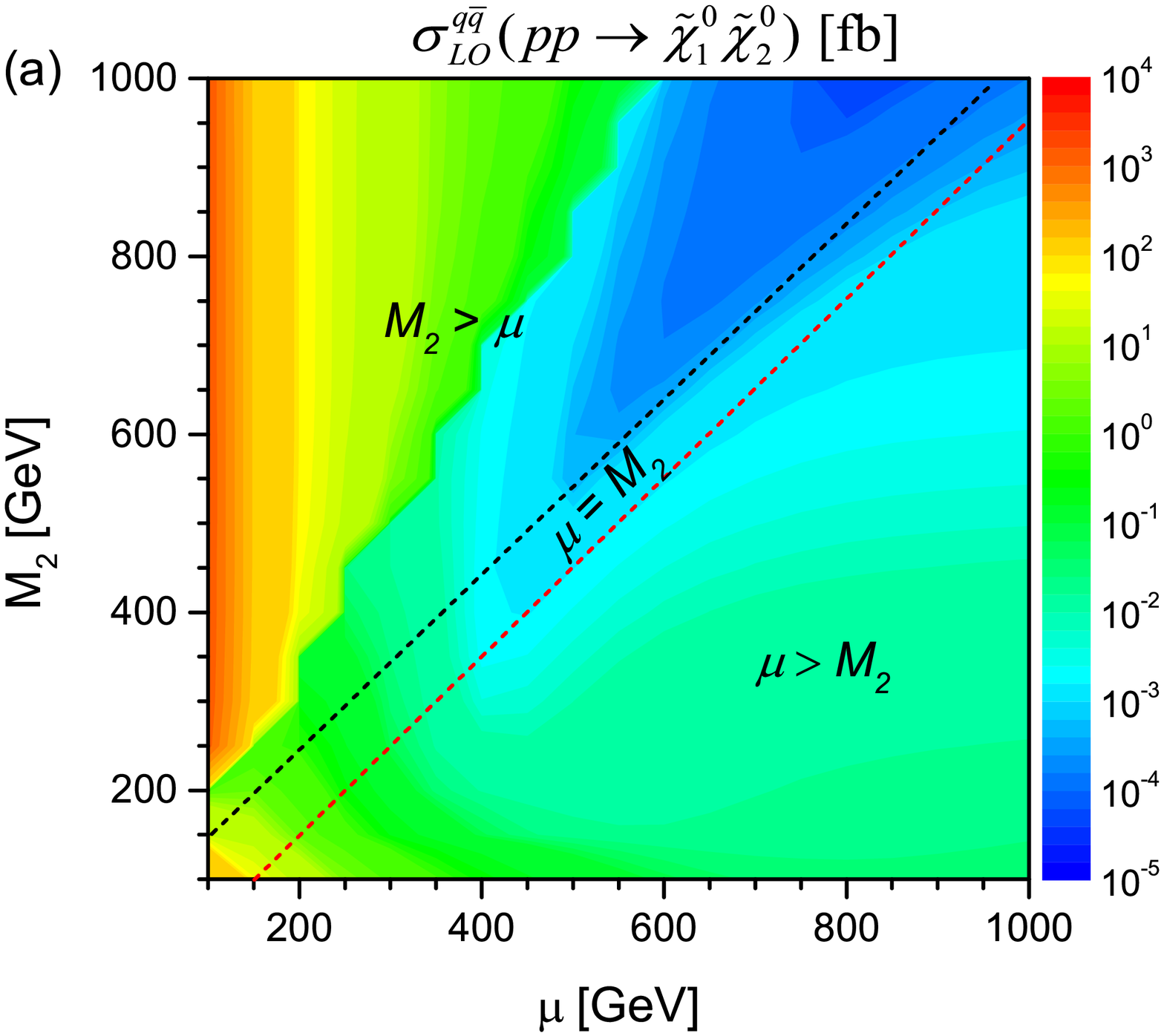}
\includegraphics[scale=0.34]{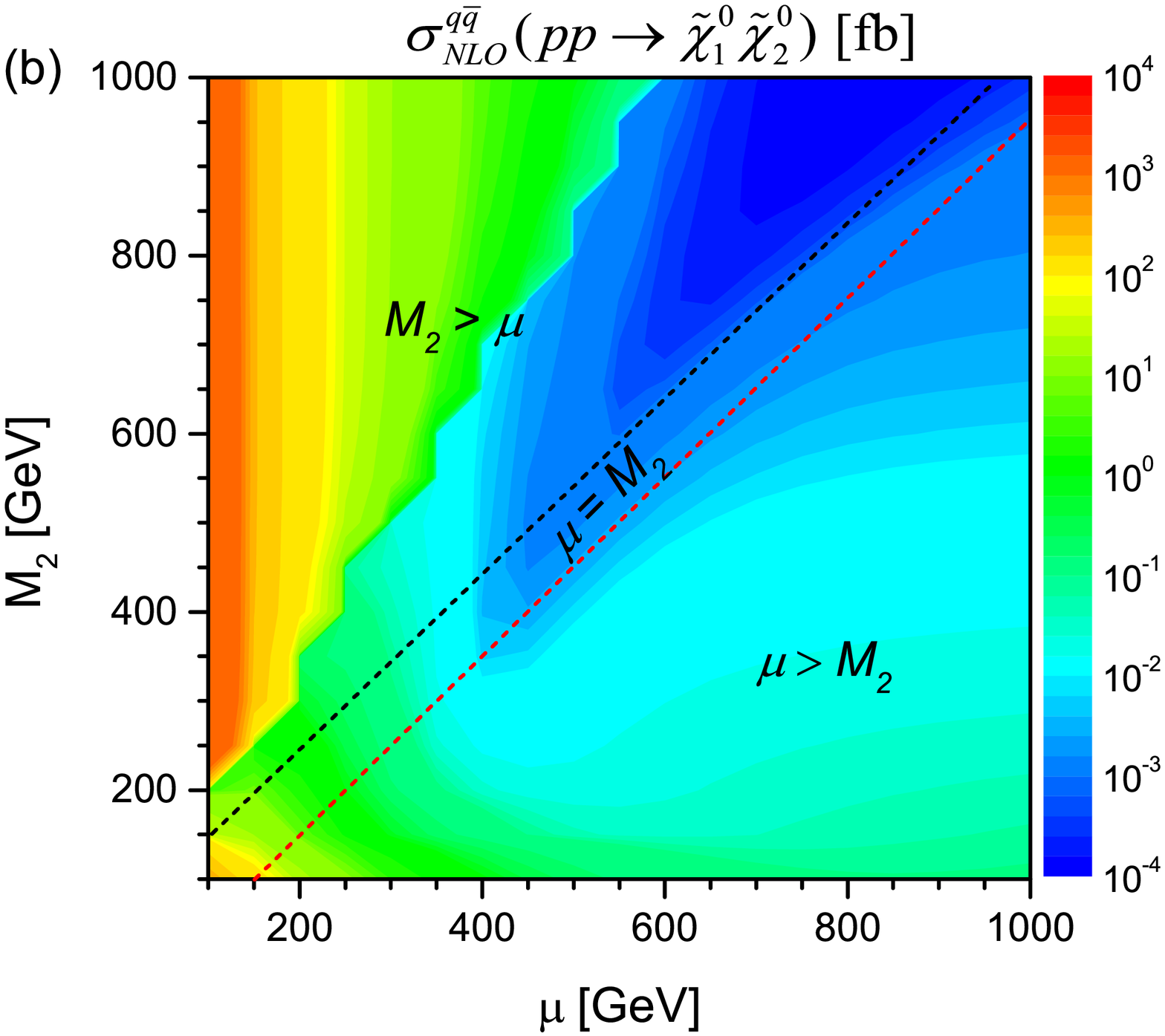}
\includegraphics[scale=0.34]{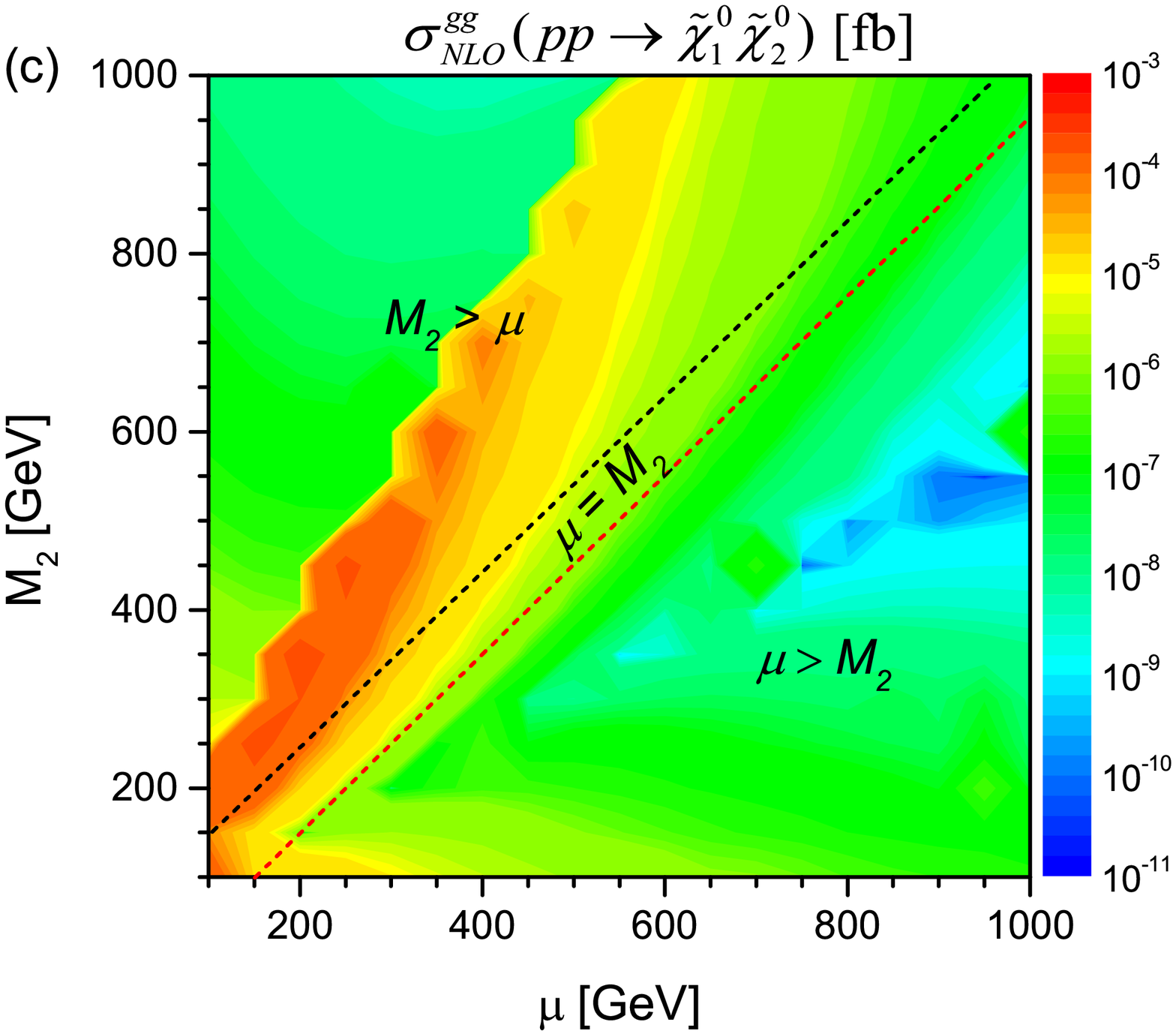}
\includegraphics[scale=0.34]{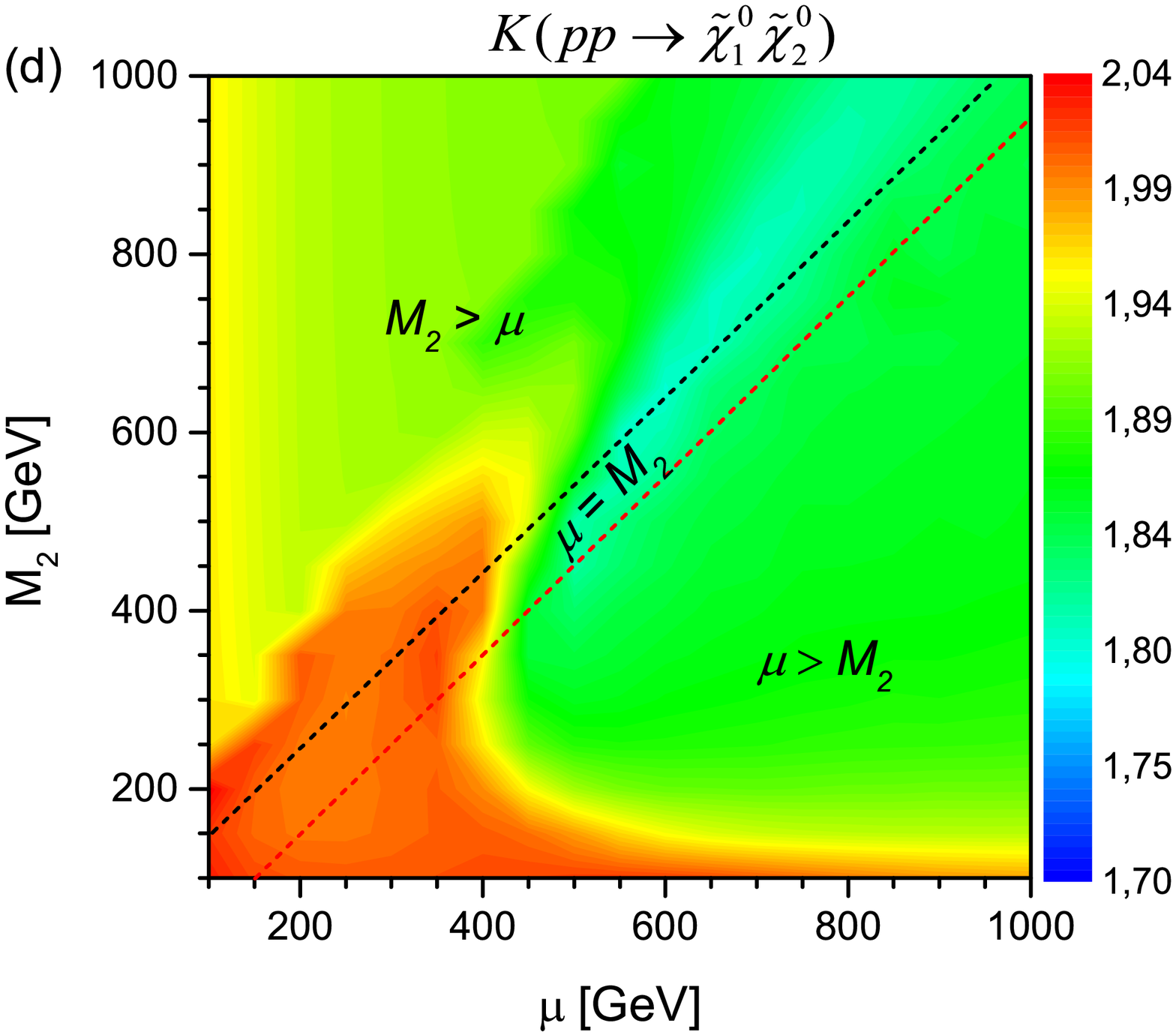}
      \end{center}
\caption{(color online). Contour plots of the total (a) LO, (b)-(c) NLO cross sections and (d) \textit{K} factor
of the process  $pp\to\widetilde{\chi}_{1}^0\widetilde{\chi}_{2}^0$ in the $M_2-\mu$ plane for $\sqrt{s}=8$ TeV,
where we take $\tanb=45$ and fix $M_1=\frac{5}{3}M_2 \tan^2\theta_W$.}
\label{fig:fig10}
\end{figure}

\begin{figure}[hpt]
    \begin{center}
\includegraphics[scale=0.34]{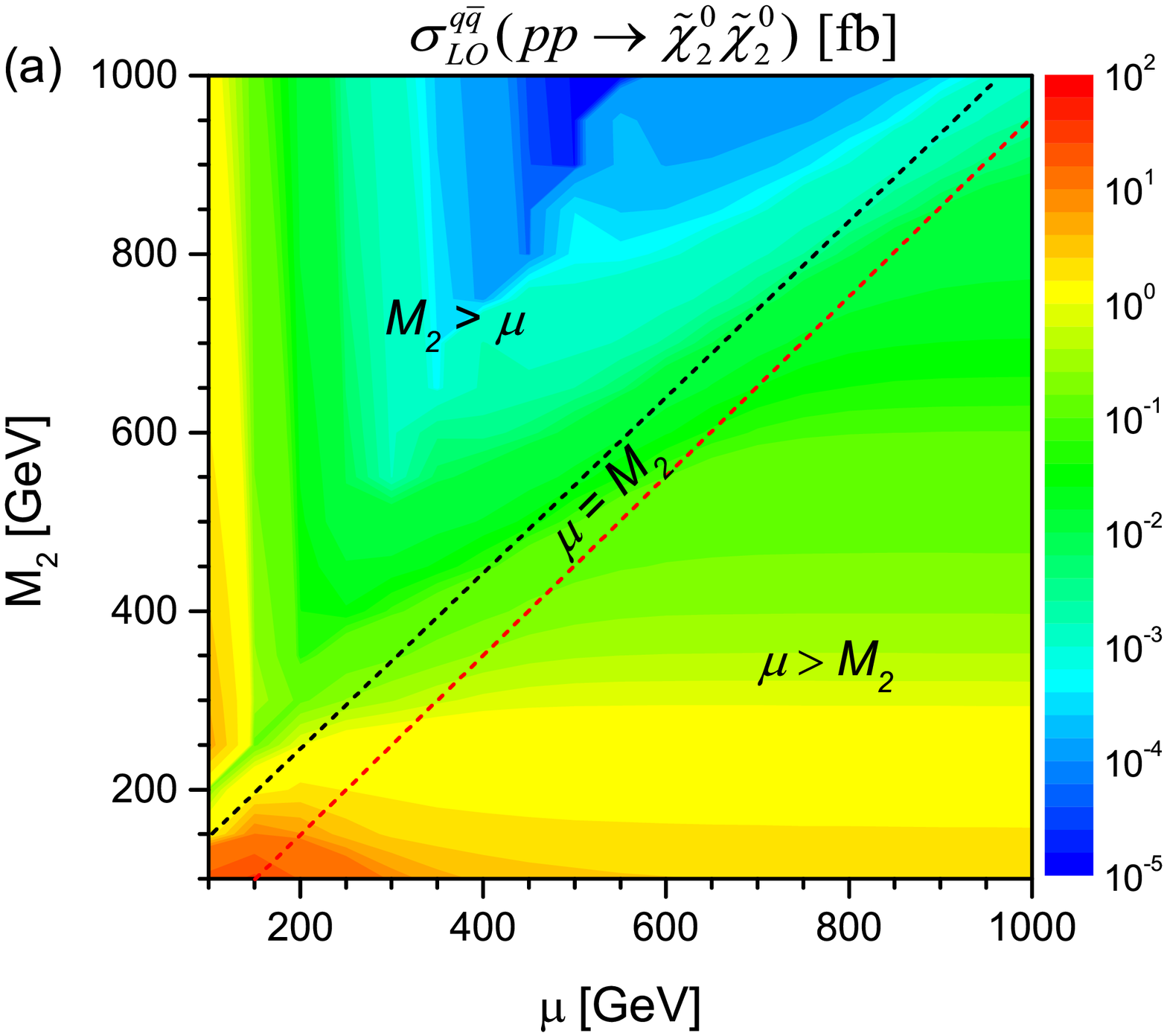}
\includegraphics[scale=0.34]{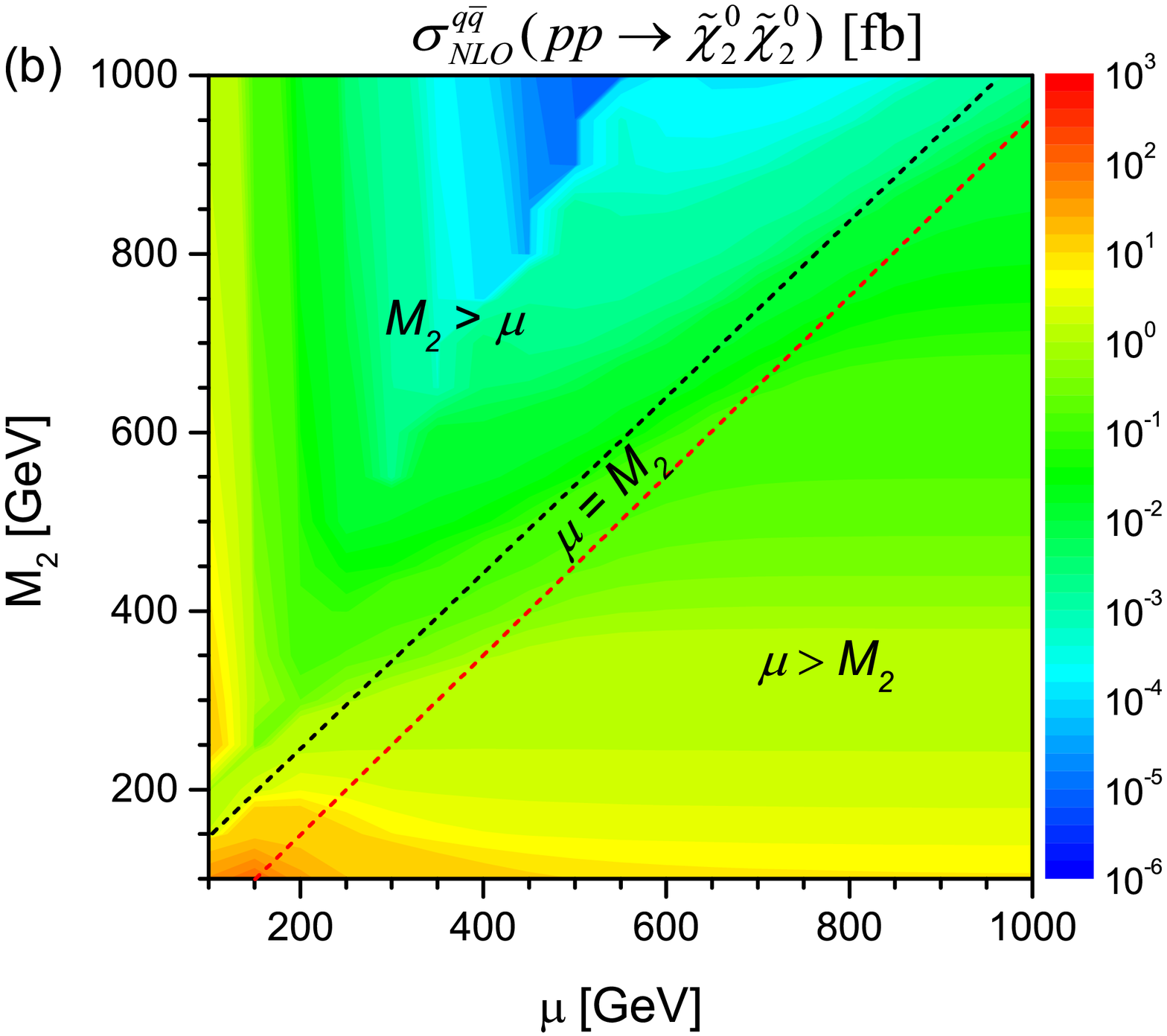}
\includegraphics[scale=0.34]{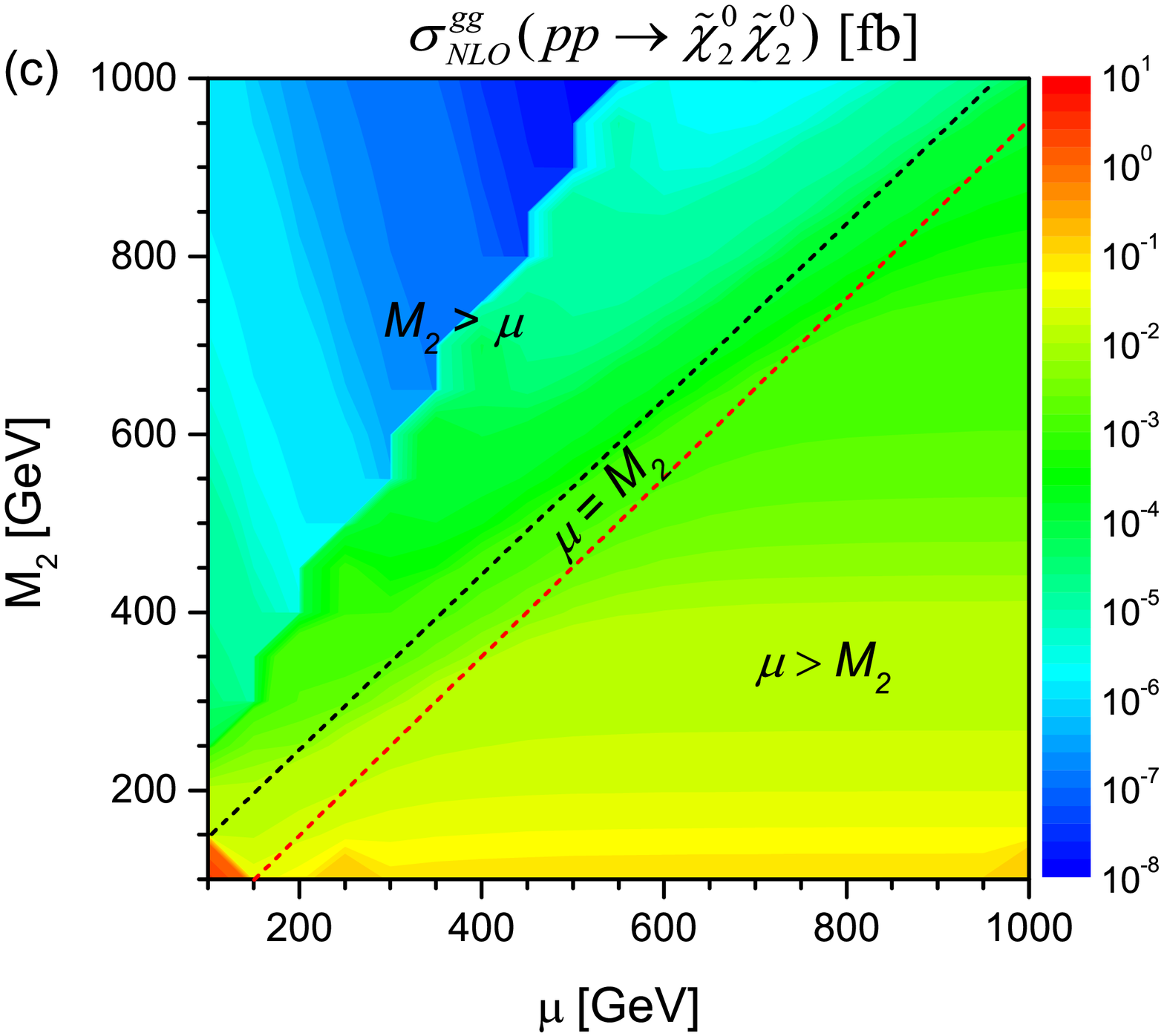}
\includegraphics[scale=0.34]{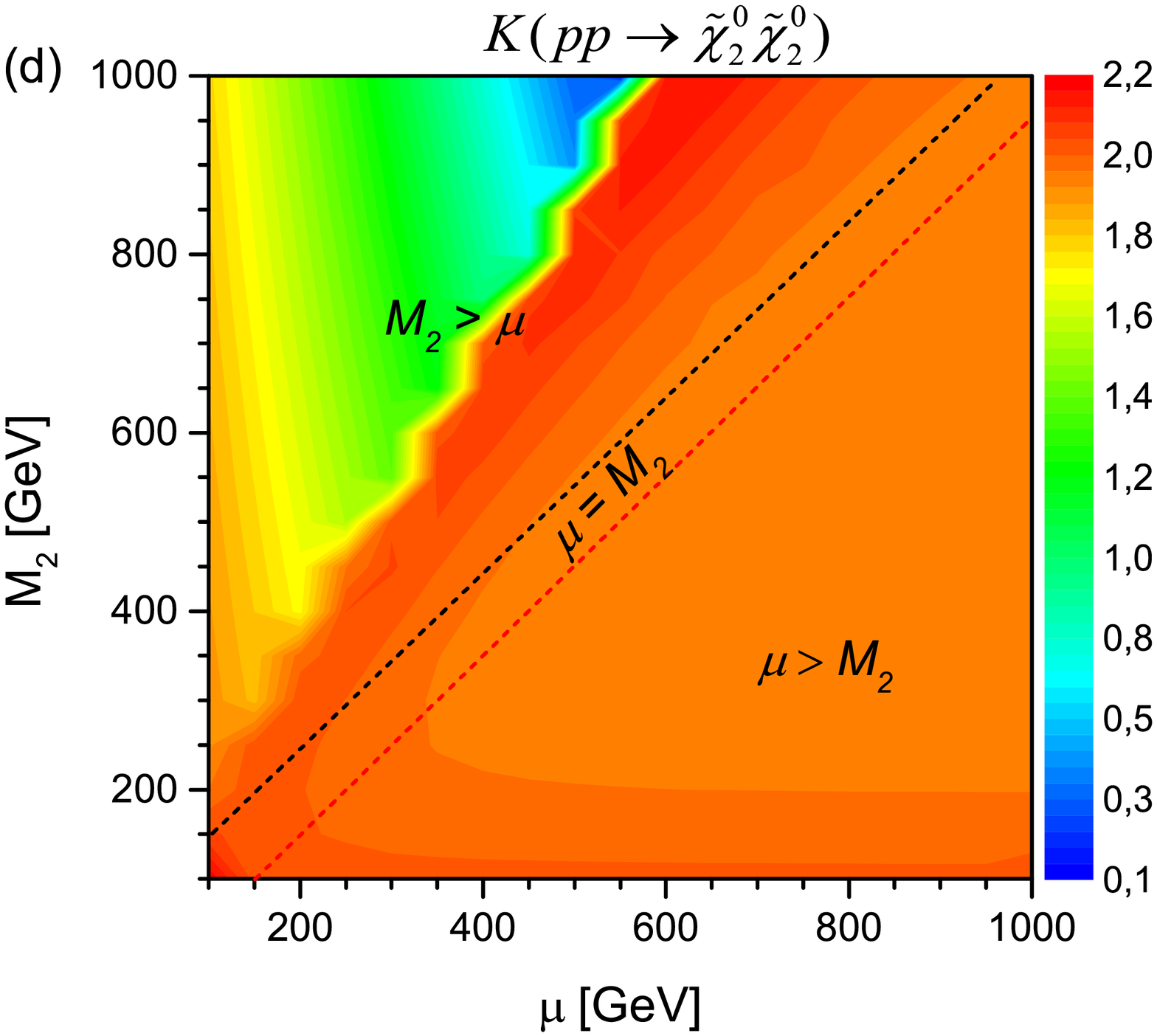}
      \end{center}
\caption{(color online). Contour plots of the total (a) LO, (b)-(c) NLO cross sections and (d) \textit{K} factor
of the process  $pp\to\widetilde{\chi}_{2}^0\widetilde{\chi}_{2}^0$ in the $M_2-\mu$ plane for $\sqrt{s}=8$ TeV,
where we take $\tanb=45$ and fix $M_1=\frac{5}{3}M_2 \tan^2\theta_W$.}
\label{fig:fig11}
\end{figure}

\begin{figure}[hpt]
    \begin{center}
\includegraphics[scale=0.32]{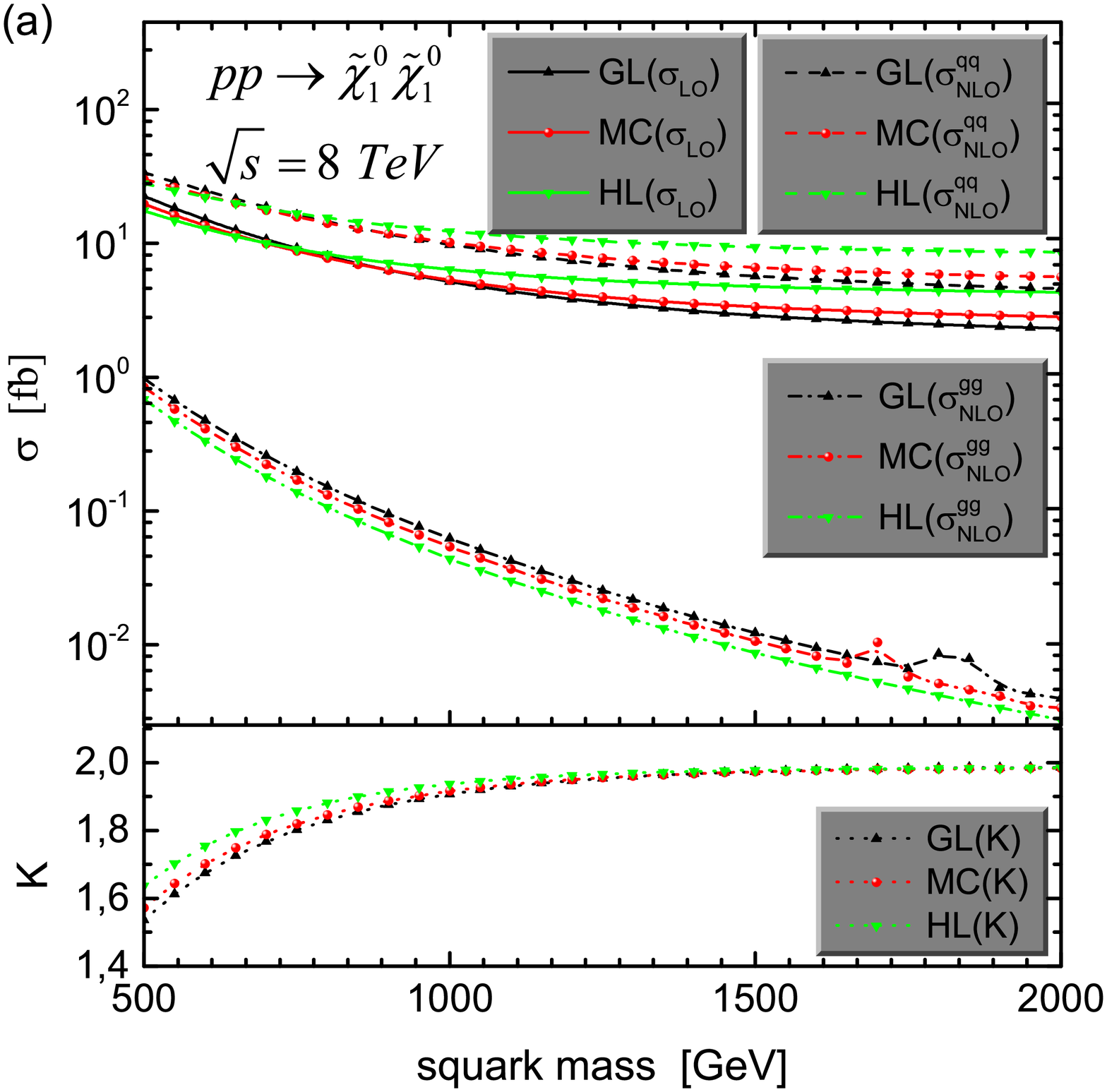}
\includegraphics[scale=0.32]{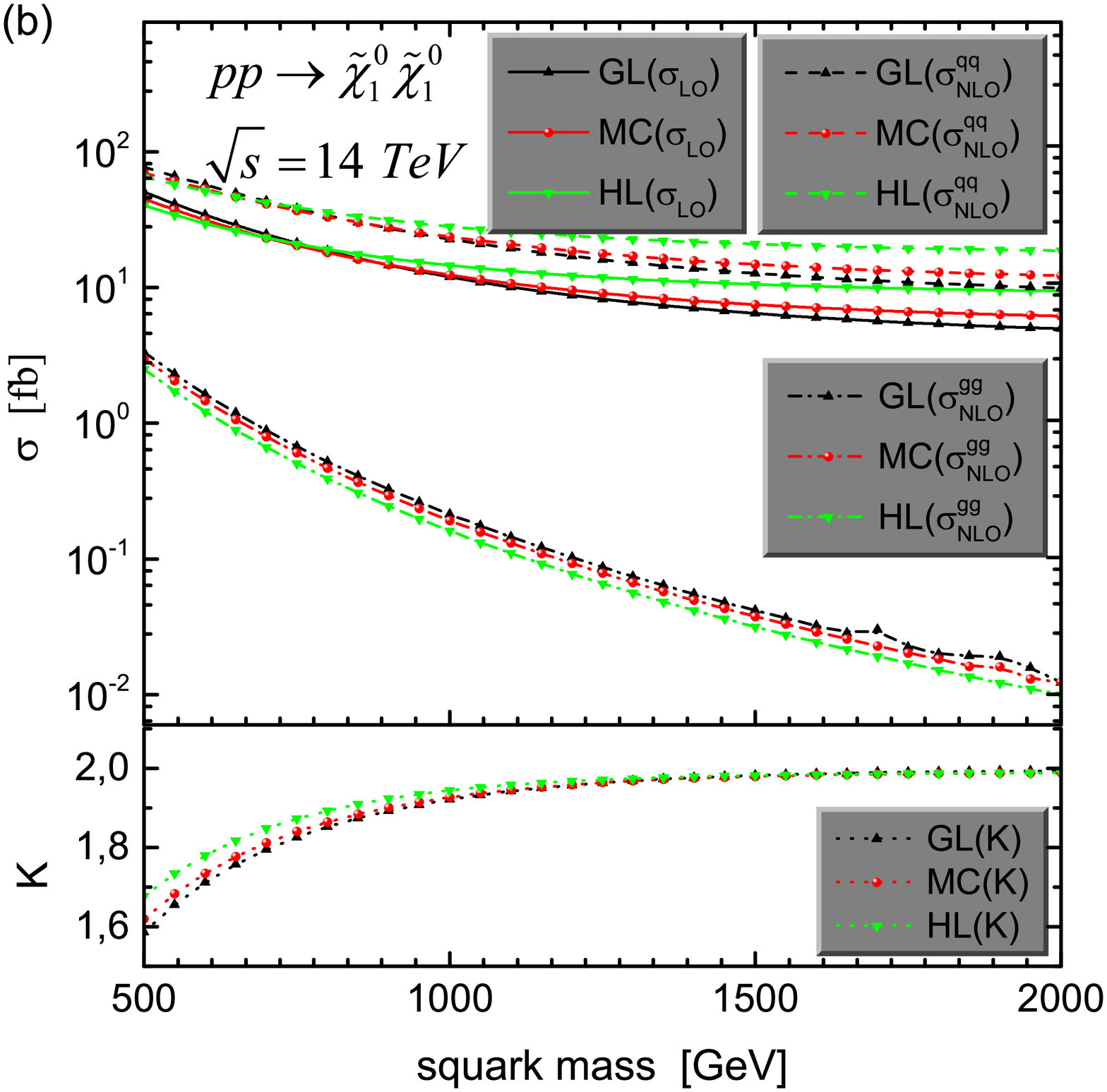}
     \end{center}
\caption{(color online). Total LO and NLO cross sections for the
process $pp\to\widetilde{\chi}_{1}^0\widetilde{\chi}_{1}^0$ depending on the squark mass at (a) $\sqrt{s}=8$ TeV
and (b) 14 TeV. The lower panels show
the \textit{K} factor, $K=(\sigma_{NLO}^{q \bar q}+\sigma_{NLO}^{\text{gg}})/\sigma_{LO}$.} \label{fig:fig12}
\end{figure}

\begin{figure}[hpt]
    \begin{center}
\includegraphics[scale=0.32]{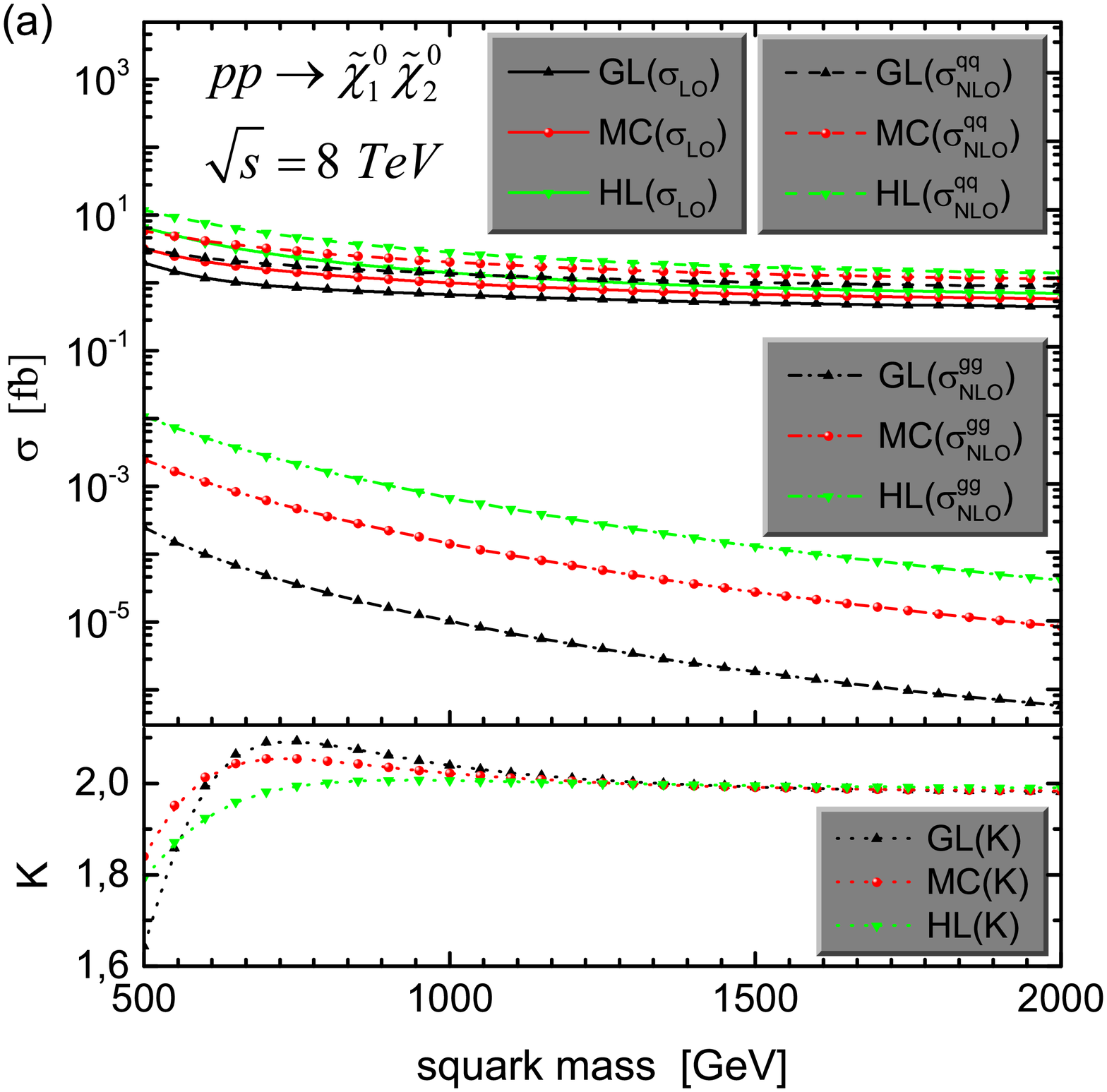}
\includegraphics[scale=0.32]{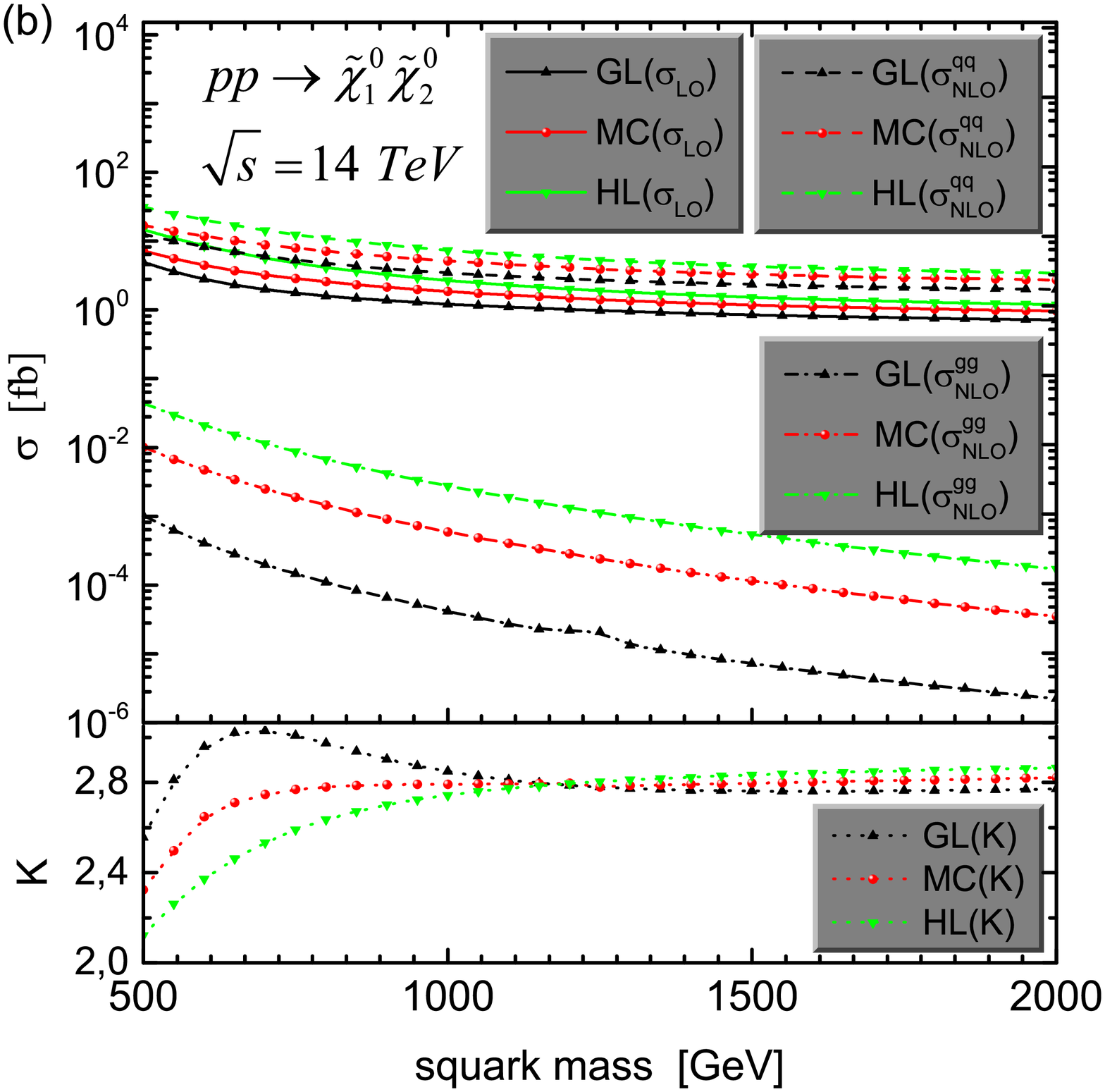}
     \end{center}
\caption{(color online). Total LO and NLO cross sections for
the process $pp\to\widetilde{\chi}_{1}^0\widetilde{\chi}_{2}^0$ depending on the squark mass
at (a) $\sqrt{s}=8$ TeV and (b) 14 TeV.
The lower panels show the \textit{K} factor, $K=(\sigma_{NLO}^{q \bar q}+\sigma_{NLO}^{\text{gg}})/\sigma_{LO}$.} \label{fig:fig13}
\end{figure}

\begin{figure*}[ht]
    \begin{center}
\includegraphics[scale=0.32]{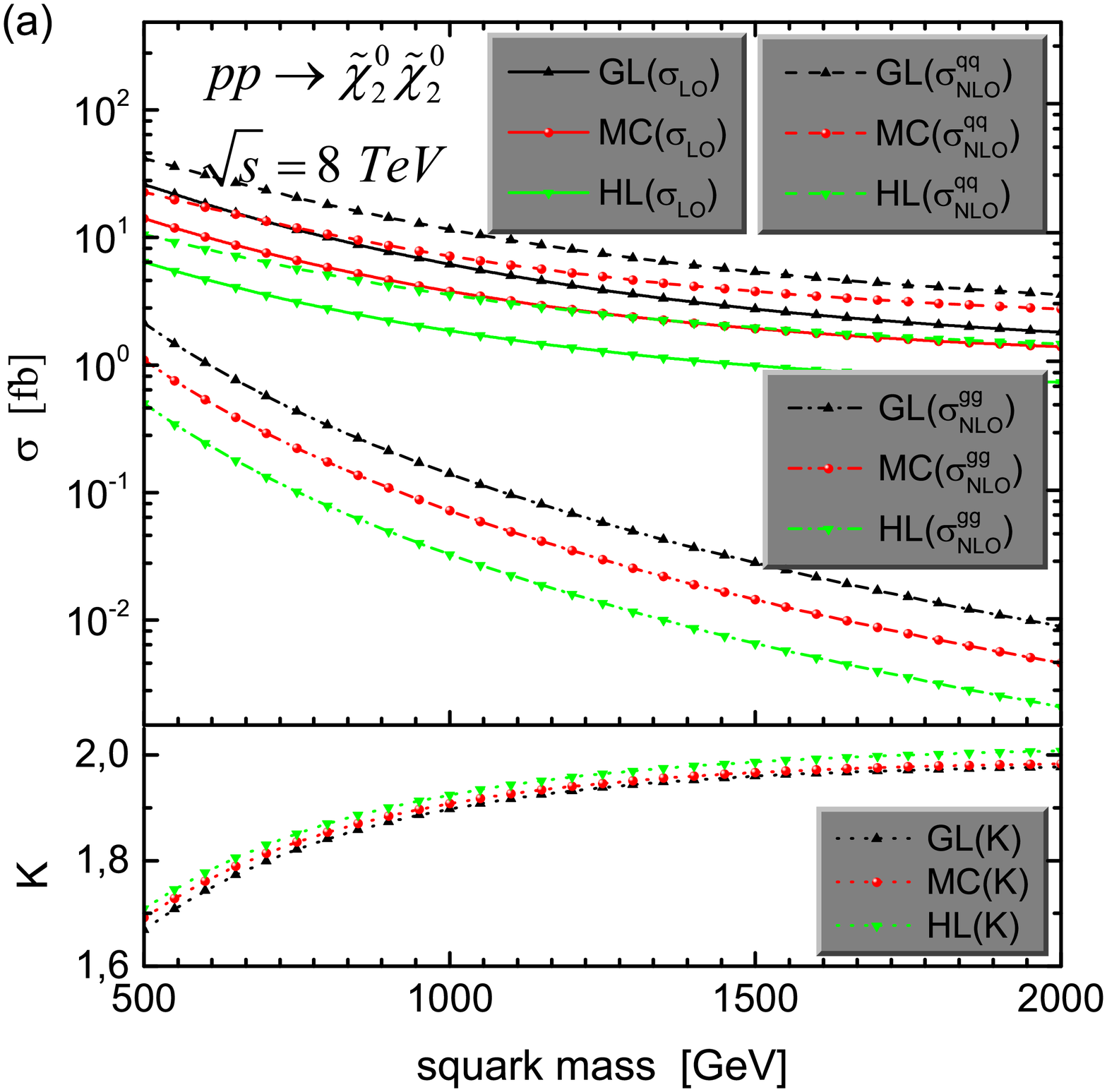}
\includegraphics[scale=0.32]{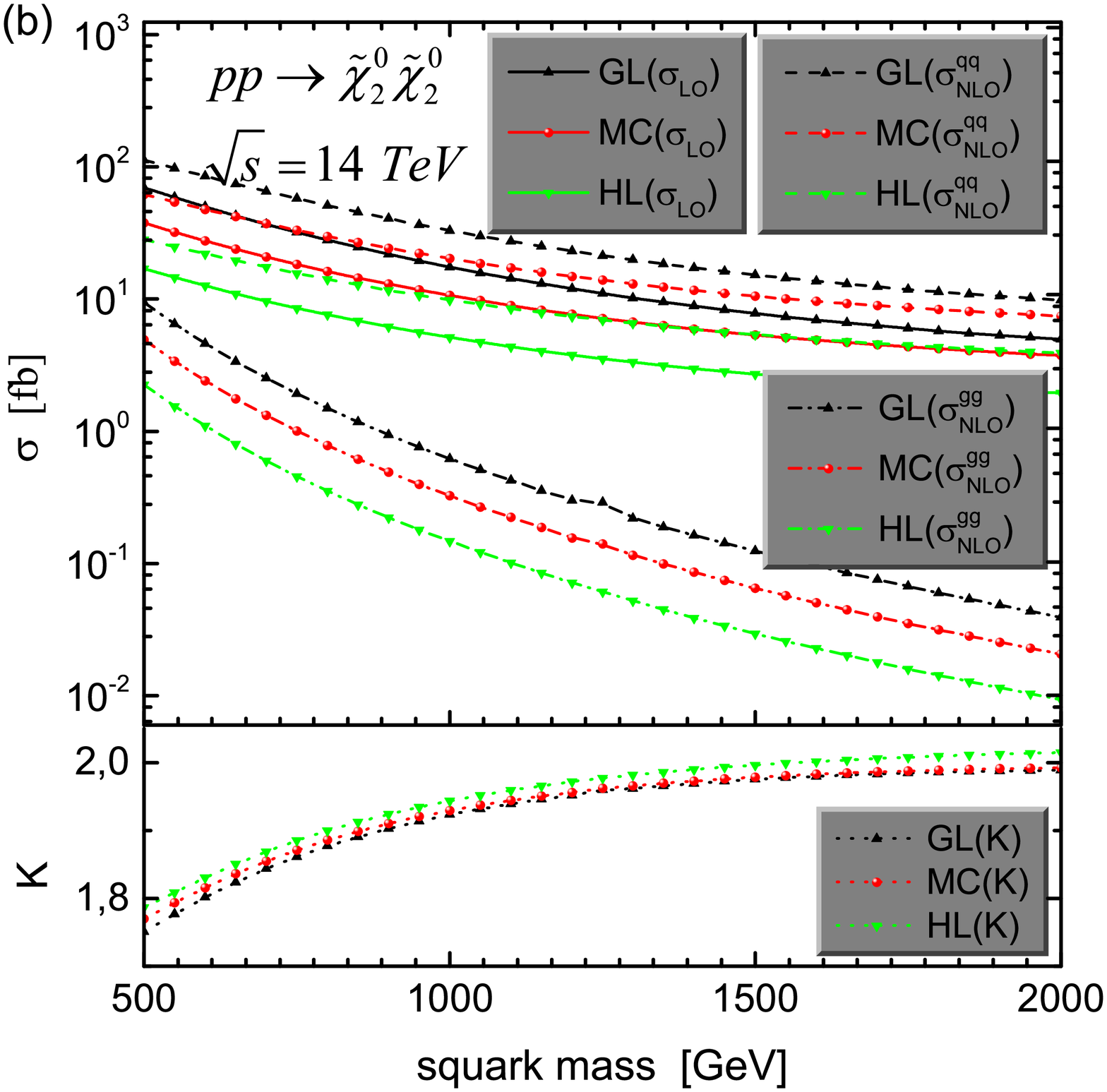}
     \end{center}
\caption{(color online). Total LO and NLO cross sections for
the process $pp\to\widetilde{\chi}_{2}^0\widetilde{\chi}_{2}^0$ depending on
the squark mass at (a) $\sqrt{s}=8$ TeV and (b) 14 TeV.
The lower panels show the \textit{K} factor, $K=(\sigma_{NLO}^{q \bar q}+\sigma_{NLO}^{\text{gg}})/\sigma_{LO}$.} \label{fig:fig14}
\end{figure*}

\begin{figure}[hpt]
    \begin{center}
\includegraphics[scale=0.324]{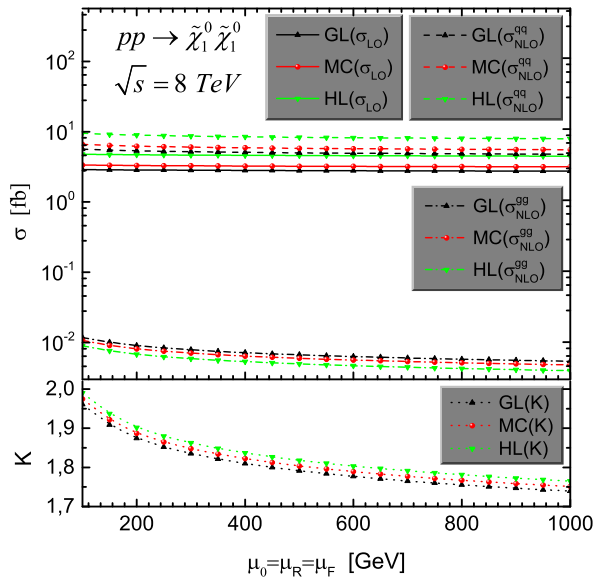}
     \end{center}
\caption{(color online). Total LO and NLO cross
sections depending on the renormalization and factorization scales for
$pp\to\widetilde{\chi}_{1}^0\widetilde{\chi}_{1}^0$ at $\sqrt{s}=8$ TeV.
The lower panel shows the \textit{K} factor, $K=(\sigma_{NLO}^{q \bar q}+\sigma_{NLO}^{\text{gg}})/\sigma_{LO}$.} \label{fig:fig15}
\end{figure}

\begin{figure}[hpt]
    \begin{center}
\includegraphics[scale=0.324]{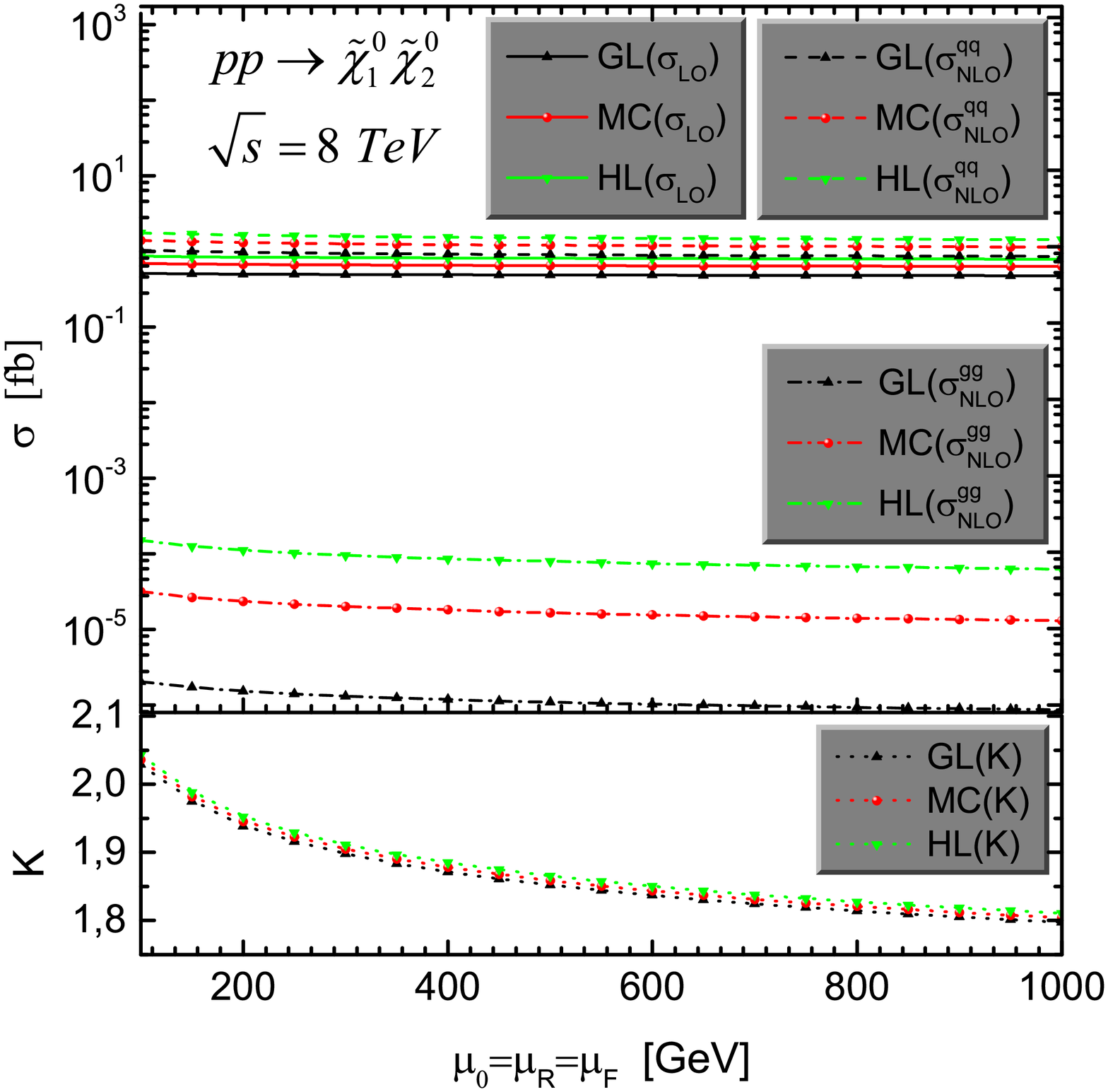}
     \end{center}
\caption{(color online). Total LO and NLO cross
sections depending on the renormalization and factorization scales for
$pp\to\widetilde{\chi}_{1}^0\widetilde{\chi}_{2}^0$ at $\sqrt{s}=8$ TeV.
The lower panel shows the \textit{K} factor, $K=(\sigma_{NLO}^{q \bar q}+\sigma_{NLO}^{\text{gg}})/\sigma_{LO}$.} \label{fig:fig16}
\end{figure}

\begin{figure}[hpt]
    \begin{center}
\includegraphics[scale=0.324]{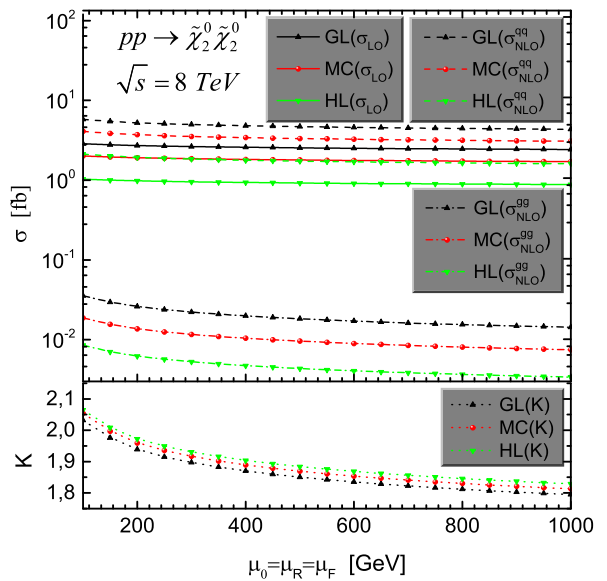}
     \end{center}
\caption{(color online). Total LO and NLO cross
sections depending on the renormalization and factorization scales for
$pp\to\widetilde{\chi}_{2}^0\widetilde{\chi}_{2}^0$ at $\sqrt{s}=8$ TeV.
The lower panel shows the \textit{K} factor, $K=(\sigma_{NLO}^{q \bar q}+\sigma_{NLO}^{\text{gg}})/\sigma_{LO}$.} \label{fig:fig17}
\end{figure}

For representative parameter points of each of the
scenarios defined in Table~\ref{tab:table1}, we have
performed numerical evaluation of the total Born cross sections $\sigma_{LO}$, the one-loop cross sections for
quark-antiquark annihilation and gluon-gluon fusion $\sigma_{NLO}^{q \bar q /\text{gg}}$, and the \textit{K} factor,
as a function of the center-of-mass energy from
Figs.~\ref{fig:fig6} through \ref{fig:fig8}, the $M_2$-$\mu$ mass plane from
Figs.~\ref{fig:fig9} through \ref{fig:fig11}, the squark mass from
Figs.~\ref{fig:fig12} through \ref{fig:fig14}, and the factorization scale from
Figs.~\ref{fig:fig15} through \ref{fig:fig17}. However, the neutralino masses and mixing matrix are not very sensitive
with respect to variation of the $\tan \beta$, so we do not illustrate any plots against it. In order to display
the numerical effect of the NLO contributions on the LO cross section, we show
the associated \textit{K} factor in the lower part of some plots. In these figures, the solid curves denote the Born cross sections,
and the dashed and dash-dotted curves represent
the one-loop cross sections for quark-antiquark annihilation and gluon-gluon fusion, respectively.
We use the following abbreviations: GL, gauginolike; HL, Higgsino-like; MC, mixture case, and
40.2.4, CMSSM 40.2.4 benchmark point. Now we present separately the following detailed analysis of these figures.

In Figs.~\ref{fig:fig6} to \ref{fig:fig8}, the dependence of the total LO cross sections, the NLO cross sections and
the \textit{K} factors on the center-of-mass energy are plotted. These plots indicate that both LO and
NLO cross sections increase smoothly and slowly with increasing the center-of-mass energy for each scenario. Moreover, the corresponding \textit{K} factors grow by about 1 percent when the center-of-mass energy increase
from 7 to 14 TeV. It implies that the \textit{K} factor is less sensitive according to varying the center-of-mass energy.
As shown in Fig.~\ref{fig:fig6}, the LO cross section of the process $pp \to \widetilde
{\chi}_{1}^{0}\widetilde{\chi}_{1}^{0}$ in the Higgsino-like scenario is roughly 41\%, 64\%, and one order of
magnitude larger than in the mixture-case scenario, gauginolike scenario, and CMSSM 40.2.4 benchmark point, respectively. The \textit{K} factors of the process $pp \to \widetilde{\chi}_{1}^{0}\widetilde{\chi}_{1}^{0}$
in our scenarios are nearly equal to each other, while they are 5\% larger than in the CMSSM 40.2.4 benchmark point. Furthermore,
one can see from Fig.~\ref{fig:fig7} that the LO cross section of the process $pp \to \widetilde
{\chi}_{1}^{0}\widetilde{\chi}_{2}^{0}$ in the Higgsino-like scenario is enhanced by about 26\%, 70\%, and two orders of magnitude relative to the mixture-case scenario, the gauginolike scenario, and CMSSM 40.2.4 benchmark point, respectively.
The \textit{K} factors for this process in our scenarios are approximately equal to each other, while they are 7\% larger than in the CMSSM 40.2.4 benchmark point. Finally, in Fig.~\ref{fig:fig8}, the LO cross section of
the process $pp \to \widetilde{\chi}_{2}^{0}\widetilde{\chi}_{2}^{0}$ in the gauginolike scenario is enhanced by around 65\%, 3 times of magnitude, and 7 times of magnitude relative to the mixture-case scenario, Higgsino-like scenario, and CMSSM 40.2.4 benchmark point, respectively. The \textit{K} factor for this process in the Higgsino-like scenario is roughly 1\%, 1\%, and 5\% larger than
in the mixture-case scenario, gauginolike scenario, and CMSSM 40.2.4 benchmark point, respectively.
\begin{table*}[ht]%
\caption{Total LO, NLO cross sections (in fb) and corresponding \textit{K} factors at center-of-mass energy $\sqrt s=$ 8 and 14 TeV for each scenario.
Here the \textit{K} factor is $K=(\sigma_{NLO}^{q \bar q}+\sigma_{NLO}^{\text{gg}})/\sigma_{LO}$.}\label{tab:table2}
\begin{ruledtabular}
\begin{tabular}{lcrrrrrrrrrrrr}
~~&~&\multicolumn{4}{c}{$pp\to\widetilde{\chi}_{1}^{0}\widetilde{\chi}_{1}^{0}$}&\multicolumn{4}{c}{$pp\to\widetilde{\chi}_{1}^{0}\widetilde{\chi}_{2}^{0}$}&\multicolumn{4}{c}{$pp\to\widetilde{\chi}_{2}^{0}\widetilde{\chi}_{2}^{0}$}\\ \cline{3-14}
scenario&$\sqrt{s}$~[TeV]&$\sigma_{LO}$&$\sigma_{NLO}^{q \bar{q}}$&$\sigma_{NLO}^{\text{gg}}$&\textit{K}~~~&$\sigma_{LO}$&$\sigma_{NLO}^{q \bar{q}}$&$\sigma_{NLO}^{\text{gg}}$&\textit{K}~~~&$\sigma_{LO}$&$\sigma_{NLO}^{q \bar{q}}$&$\sigma_{NLO}^{\text{gg}}$&\textit{K}~~~\\
 \hline
\multirow{2}*{Higgsino-like}&8&4.76&9.40&0.009&1.98&0.85&1.70&1.3$\cdot10^{-4}$&2.00&0.99&1.96&0.007&1.99\\
                            &14~~&10.54& 20.87&0.032 &1.98&1.95&3.90&5.4$\cdot10^{-4}$&2.00&2.71&5.38&0.029&2.00\\
\noalign{\smallskip}
\multirow{2}*{Gauginolike} &8&2.89&5.70&0.012&1.97&0.51&1.01&1.8$\cdot10^{-6}$&1.99&2.75&5.36&0.028&1.96\\
                            &14~~&6.45& 12.74&0.042&1.98& 1.11&2.22&7.2$\cdot10^{-6}$&2.00&7.78&15.25&0.124&1.98\\
\noalign{\smallskip}
\multirow{2}*{Mixture case} &8&3.35&6.60&0.011&1.97&0.68&1.36&2.7$\cdot10^{-5}$&1.99&1.93&3.77&0.014&1.97\\
                            &14~~&7.49& 14.79&0.038&1.98& 1.53&3.05&1.2$\cdot10^{-4}$&1.99&5.35&10.53&0.065&1.98\\
\noalign{\smallskip}
\multirow{2}*{CMSSM 40.2.4} &8&0.31&0.58&0.005&1.88&0.01&0.02&1.2$\cdot10^{-9}$&1.87&0.27&0.51&0.007&1.92\\
                            &14~~&1.15& 2.15&0.032&1.90&0.04&0.08&1.5$\cdot10^{-7}$&1.88&1.63&3.11&0.071&1.95\\
\end{tabular}
\end{ruledtabular}
\end{table*}

We document a numerical survey over our scenarios and the CMSSM 40.2.4 benchmark point for
LHC center-of-mass energies of 8 and 14 TeV in Table~\ref{tab:table2}.
One can deduce from above analysis and this table that the
total LO and NLO cross section of the process $pp \to
\widetilde{\chi}_{1}^{0}\widetilde{\chi}_{1}^{0}$ in the
Higgsino-like scenario is usually larger than others.
The LO (NLO) cross section of the process $pp \to \widetilde
{\chi}_{1}^{0}\widetilde{\chi}_{1}^{0}$ in the Higgsino-like
scenario appears in the range of 3.9 to 10.5 ($\sigma_{\text{NLO}}=$ 7.6 to 20.9) fb, resulting in \textit{K} factor of about $K=$ 1.98. Furthermore, for process $pp \to \widetilde
{\chi}_{2}^{0}\widetilde{\chi}_{2}^{0}$ in the gauginolike
scenario, the cross section appears in the range of 2.07 to 7.8 ($\sigma_{\text{NLO}}=$ 4.05 to 15.4) fb, resulting in \textit{K} factor of $K=$ 1.96 to 1.98 and should be observable at LHC.
The quark-antiquark annihilation yields larger NLO cross section than gluon-gluon fusion for each scenario.
The sizes of the NLO cross sections are at a visible level of $10^{-1}$ fb for gg fusion while $10^{1}$ fb
for $q \bar{q}$ annihilation. Particularly, for process $\text{gg} \to \widetilde
{\chi}_{2}^{0}\widetilde{\chi}_{2}^{0}$ in the gauginolike
scenario, the cross section reaches a value of 0.124 fb at $\sqrt s =$ 14 TeV. Moreover, as one sees from Table~\ref{tab:table2}, the NLO contributions for neutralino pair production are so significant that the \textit{K} factor yields around $K\sim$2.
One notes that the associated \textit{K} factors barely change between our scenarios according to the dependence on
the center-of-mass energy. This behavior between \textit{K} factors is shown to be ordered
as HL(\textit{K})$\sim$GL(\textit{K})$\sim$MC(\textit{K})$>$CMSSM(\textit{K}).

The neutralino/chargino masses and mixing matrices depend on the
$M_{2}$ and $\mu$ mass parameters so significantly that the interesting information can be obtained from the dependence of the cross section on these parameters. Correspondingly,
we investigate the effect of these parameters on the LO, NLO cross sections and the \textit{K} factors of
the relevant process in $M_{2}$-$\mu$ mass plane with varying these parameters
in the range from 100 to 1000 GeV in steps of 50 GeV at $\sqrt{s}=8$ TeV for $\tanb=$ 45, as illustrated in Figs.~\ref{fig:fig9} through \ref{fig:fig11}.
In these plots, the region below the red dashed line corresponds to gauginolike ($\mu>M_2$), the region above the black dashed line corresponds to Higgsino-like ($M_2>\mu$), and the region between the two
dashed lines corresponds to mixture case ($\mu=M_2$).
One sees that the LO and NLO cross sections increase with
decreasing  $M_{2}$ and any value of $\mu$ for $pp \to \neutralino_{1}\neutralino_{1}$
and $pp \to \neutralino_{2}\neutralino_{2}$, whereas decreasing  $\mu$ and any value of $M_{2}$
for  $pp \to \neutralino_{1}\neutralino_{2}$. In particular, cross section reaches
maximal values in the region $M_{2}\lesssim 400$ GeV for  $pp \to \neutralino_{1}\neutralino_{1}$
and $\neutralino_{2}\neutralino_{2}$, and $\mu\lesssim 500$ GeV for  $pp \to \neutralino_{1}\neutralino_{2}$
into the scan region. From these results one can conclude that pure gaugino couplings dominate
in the case of same type of neutralinos $i=j$, whereas pure Higgsino couplings enhance in the case of different
type of neutralinos $i\neq j$ for $pp \to \neutralino_{i}\neutralino_{j}$. The \textit{K} factors have mostly
the values in the range between 2.3 to 1.8 for $pp \to \neutralino_{1}\neutralino_{1}$ [shown in Fig.~\ref{fig:fig9}{\color{red}(d)}], 2.0
to 1.8 for $pp \to \neutralino_{1}\neutralino_{2}$ [shown in Fig.~\ref{fig:fig10}{\color{red}(d)}] and 2.2 to 1.9
for $pp \to \neutralino_{2}\neutralino_{2}$ [shown in Fig.~\ref{fig:fig11}{\color{red}(d)}] in the scan region.
The maximum values of the \textit{K} factor are obtained in the region $\mu \lesssim 500$ GeV and
400 $\lesssim M_2 \lesssim$ 1000 GeV for processes $pp \to \neutralino_{1}\neutralino_{1}$, $\mu \lesssim 500$ GeV
and $M_2 \lesssim$ 500 GeV for processes $pp \to \neutralino_{1}\neutralino_{2}$ and  $\mu>M_{2}$ for
process $pp \to \neutralino_{2}\neutralino_{2}$. For example the \textit{K} factor increases from 1.45 to 2.22
for $pp \to \neutralino_{1}\neutralino_{1}$, whereas decreases from 2.01 to 1.93
for $pp \to \neutralino_{1}\neutralino_{2}$ and 1.99 to 1.44 for $pp \to \neutralino_{2}\neutralino_{2}$
with the increment of $M_2$ from 100 to 1000 GeV at $\mu=$ 200 GeV. What's more, when the parameter $\mu$ varies
from 100 to 1000 GeV for $M_2=$ 200 GeV, the \textit{K} factor decreases from 2.02 to 1.94
for $pp \to \widetilde{\chi}_{1}^{0}\widetilde{\chi}_{1}^{0}$, from 2.03 to 1.90
for $pp \to \widetilde{\chi}_{1}^{0}\widetilde{\chi}_{2}^{0}$ and from 1.94 to 1.93
for $pp \to \widetilde{\chi}_{2}^{0}\widetilde{\chi}_{2}^{0}$. As a consequence,
it is clearly visible that the \textit{K} factor strongly depends on the parameters $M_{2}$ and $\mu$.

\newpage
In Figs.~\ref{fig:fig12} to \ref{fig:fig14}, we show the dependence
of the total LO, NLO cross sections and the \textit{K} factors on the squark mass for each scenario
at $\sqrt{s}=$ 8 and 14 TeV. We vary the squark mass from 500 to 2000 GeV. Here, there arise
the same dominant scenarios as ones in the center-of-mass energy dependence of the cross sections.
The LO and NLO cross sections for both $q \bar{q}$ annihilation and gg fusion are mainly determined by
the squark mass. These decrease with the increment of the squark mass from 500 to 2000 GeV. When the squark mass grows by a factor of 4, the NLO cross sections are reduced by around one and two orders of magnitude
for $q \bar{q}$ annihilation and gg fusion, respectively. We can see that the \textit{K} factor is sensitive
according to increment of the squark mass.
When the squark mass runs
from 500 to 2000 GeV at center-of-mass energy 8 TeV (14 TeV), the \textit{K} factor increases
from 1.54 to 1.99 (1.59 to 1.99) in the gauginolike scenario,
from 1.57 to 1.98 (1.62 to 1.99) in the mixture-case scenario, and from 1.64 to 1.98 (1.68 to 1.99)
in the Higgsino-like scenario for the process $pp \to\widetilde{\chi}_{1}^{0}\widetilde{\chi}_{1}^{0}$ as shown
in Figs.~\ref{fig:fig12}{\color{red}(a)}-\ref{fig:fig12}{\color{red}(b)}.
Furthermore, the \textit{K} factor for the process $pp \to\widetilde{\chi}_{1}^{0}\widetilde{\chi}_{2}^{0}$ increases from 1.64 to 1.98 (2.56 to 2.77) in the gauginolike scenario,
\begin{table*}[ht]%
\caption{Total LO, NLO cross sections (in fb) and corresponding \textit{K} factors as a function of the squark mass at center-of-mass energy $\sqrt s=$ 8 and 14 TeV
for each scenario. Here the \textit{K} factor is $K=(\sigma_{NLO}^{q \bar q}+\sigma_{NLO}^{\text{gg}})/\sigma_{LO}$.}\label{tab:table3}
\begin{ruledtabular}
\begin{tabular}{lccrrrcrrccrrrc}
&&&\multicolumn{4}{c}{$pp\to\widetilde{\chi}_{1}^{0}\widetilde{\chi}_{1}^{0}$}&\multicolumn{4}{c}{$pp\to\widetilde{\chi}_{1}^{0}\widetilde{\chi}_{2}^{0}$}&\multicolumn{4}{c}{$pp\to\widetilde{\chi}_{2}^{0}\widetilde{\chi}_{2}^{0}$}\\ \cline{4-15}
&$\sqrt{s}$~(TeV)&$m_{\widetilde{q}}$~(TeV)&$\sigma_{LO}$~&$\sigma_{NLO}^{q \bar{q}}$&$\sigma_{NLO}^{\text{gg}}$&\textit{K}~&$\sigma_{LO}$~&$\sigma_{NLO}^{q \bar{q}}$&$\sigma_{NLO}^{\text{gg}}$&\textit{K}~&$\sigma_{LO}$~&$\sigma_{NLO}^{q \bar{q}}$&$\sigma_{NLO}^{\text{gg}}$&\textit{K}~\\
 \hline
\multirow{4}*{HL}&\multirow{2}*{8} &1&6.37&12.30&0.044&1.94&1.40&2.82&6.7$\cdot10^{-4}$&2.01&1.85&3.53&0.033&1.92\\
                                   &&2&4.32&8.57&0.003&1.98&0.70&1.39&4.1$\cdot10^{-5}$&1.99&0.73&1.46&0.002&2.01\\
&\multirow{2}*{14}&1&14.55& 28.13&0.16 &1.94&2.68&7.35&2.8$\cdot10^{-3}$&2.74&5.13&9.82&0.15&1.94\\
                &&2&9.41& 18.71&0.001 &1.99&1.20&3.43&1.7$\cdot10^{-4}$&2.87&1.95&3.91&0.009&2.01\\
\hline
\noalign{\smallskip}
\multirow{4}*{GL}&\multirow{2}*{8} &1&5.17&9.79&0.062&1.91&0.67&1.37&1.0$\cdot10^{-5}$&2.04&6.15&11.54&0.14&1.90\\
                                   &&2&2.31&4.59&0.004&1.99&0.45&0.88&5.7$\cdot10^{-7}$&1.98&1.81&3.57&0.009&1.98\\
&\multirow{2}*{14}&1&11.94& 22.73&0.21 &1.92&1.21&3.46&4.1$\cdot10^{-5}$&2.85&17.47&32.98&0.62&1.92\\
                &&2&4.97& 9.90&0.012 &1.99&0.72&1.99&2.2$\cdot10^{-6}$&2.77&4.95&9.81&0.039&1.99\\
\hline
\noalign{\smallskip}
\multirow{4}*{MC}&\multirow{2}*{8} &1&5.33&10.16&0.054&1.92&1.00&2.02&1.4$\cdot10^{-4}$&2.02&3.78&7.15&0.072&1.91\\
                                   &&2&2.83&5.61&0.003&1.98&0.58&1.14&8.5$\cdot10^{-6}$&1.98&1.39&2.75&0.005&1.98\\
&\multirow{2}*{14}&1&12.35& 23.62&0.19 &1.93&1.85&5.18&5.9$\cdot10^{-4}$&2.79&10.64&20.21&0.32&1.93\\
                &&2&6.15& 12.23&0.012 &1.99&0.97&2.73&3.5$\cdot10^{-5}$&2.82&3.74&7.43&0.021&1.99\\
\end{tabular}
\end{ruledtabular}
\end{table*}
from 1.84 to 1.98 (2.32 to 2.82) in the mixture-case scenario, and
from 1.79 to 1.99 (2.12 to 2.87) in the Higgsino-like scenario for center-of-mass energy 8 TeV (14 TeV) as shown
in Figs.~\ref{fig:fig13}{\color{red}(a)}-\ref{fig:fig13}{\color{red}(b)}. Finally, the \textit{K} factor
for the process $pp \to\widetilde{\chi}_{2}^{0}\widetilde{\chi}_{2}^{0}$ increases from 1.67 to 1.98 (1.75 to 1.99)
in the gauginolike scenario, from 1.69 to 1.98 (1.77 to 1.99) in the mixture-case scenario, and
from 1.71 to 2.01 (1.79 to 2.01) in the Higgsino-like scenario at center-of-mass energy 8 TeV (14 TeV) as illustrated
in Figs.~\ref{fig:fig14}{\color{red}(a)}-\ref{fig:fig14}{\color{red}(b)}.
With a view to make easy precise comparisons with the experimental results,
we list in Table~\ref{tab:table3} the numerical results of the LO, NLO cross sections
and \textit{K} factors for the squark mass 1 and 2 TeV at $\sqrt{s}=$ 8 and 14 TeV.
The results show that the NLO corrections increase the corresponding LO cross sections when squark mass varies from 500 to 2000 GeV.
\par
Finally, the total LO, NLO cross sections and the \textit{K} factors for the process $pp \to
\widetilde {\chi}_{i}^{0}\widetilde{\chi}_{j}^{0}$ versus
the factorization and renormalization scale $\mu_0$ in the range from 100 to 1000 GeV at $\sqrt{s}=8$ TeV are depicted
in Figs.~\ref{fig:fig15} through \ref{fig:fig17}. These figures demonstrate the same dominant scenarios
as ones in the dependence of the cross sections on the center-of-mass energy.
From these figures we can also see that both LO and one-loop cross sections decrease slightly when the scale $\mu_0$
goes up from 100 to 1000 GeV for each scenario. One can remark that the LO cross sections are nearly independent of the scale $\mu_0$. That is since the neutralino pair production process at
Born-level contains only pure electroweak channels where there is not the renormalization scale dependence at the this level, and the energy scale dependence is only due to the PDFs being connected to the factorization scale. The corresponding \textit{K} factors, on the other hand, decrease
by about 11 percent when the scale $\mu_0$ varies from 100 to 1000 GeV for each scenario.
Figure~\ref{fig:fig15} shows that the \textit{K} factor for $pp\to\widetilde {\chi}_{1}^{0}\widetilde{\chi}_{1}^{0}$ decreases from 1.99 to 1.76,  1.96 to 1.74, and 1.97 to 1.75 in the Higgsino-like scenario, gauginolike scenario, and
mixture-case scenario when increasing the scale $\mu_0$ from 100 to 1000 GeV, respectively.
Figure~\ref{fig:fig16} shows that the \textit{K} factor for $pp\to\widetilde {\chi}_{1}^{0}\widetilde{\chi}_{2}^{0}$ decreases
from 2.04 to 1.81,  2.03 to 1.80, and 2.04 to 1.80 in the Higgsino-like scenario, gauginolike scenario, and
mixture-case scenario when increasing the scale $\mu_0$ from 100 to 1000 GeV, respectively.
Figure~\ref{fig:fig17} displays that the \textit{K} factor for $pp\to\widetilde {\chi}_{2}^{0}\widetilde{\chi}_{2}^{0}$
decreases from 2.06 to 1.83,  2.03 to 1.79, and 2.05 to 1.81 in the Higgsino-like scenario, gauginolike scenario, and
mixture-case scenario when increasing the scale $\mu_0$ from 100 to 1000 GeV, respectively. These results show that
the \textit{K} factors are mostly sensitive to the scale $\mu_0$.

It should be noted that analysis of our calculations is not directly dependent on the mass of the Higgs.
The \textit{K} factor is not sensitive to the mass of the
Higgs. We have a figure with full spectrum. However, we also have
alternative scenarios and have compared them with the CMSSM benchmark point. As is
seen in the figures, the cross sections calculated in the alternative
scenarios are more dominant according to the CMSSM 40.2.4 benchmark point.

We can see that one-loop contributions are
positive and essentially increase the LO
cross sections. Additionally, the curves in the figures display that the one-loop cross sections for gg fusion have a larger incline than the $q\bar{q}$ annihilation; the effect being primarily on account of the behavior of the gluon PDFs.
It can be also seen that the one-loop cross sections for the $q\bar{q}$ annihilation always are larger than
the LO cross sections, while one-loop cross sections for the gg fusion are less than these. However, note that the gg fusion contribution can be comparable to the $q\bar{q}$ one in the low $\tanb$ regime.

\section{Conclusion}\label{sec:conclusion}
In this work we have computed one-loop contributions for the neutralino pair production processes
via quark-antiquark annihilation and gluon-gluon fusion in proton-proton collisions at
the LHC. We have investigated numerically the effects of the center-of-mass energy,
the $M_2$-$\mu$ mass plane, the squark mass, the factorization and renormalization scales on the total Born,
NLO cross sections and \textit{K} factor for the CMSSM, and three different scenarios called the gauginolike, Higgsino-like, and mixture case.

The numerical results show that the NLO corrections increase the Born cross sections. The one-loop contributions of the process $q\bar{q}$ annihilation are significant for
the experimental and phenomenological works in connection with the neutralino pair productions at the LHC
and the future colliders. The \textit{K} factor varies in the range
from 1.96 to 2.00 when center-of-mass energy goes from 7 to 14 TeV.
Our scenarios dominate over the CMSSM 40.2.4 benchmark scenario for each process.
It is clear that the strong dependence of the cross sections and \textit{K} factor
on the parameters $M_{2}$ and $\mu$ is remarkable. From the discussed results in the $M_2$-$\mu$ mass plane,
we can conclude that pure gaugino couplings dominate in the case $i=j$, whereas pure Higgsino couplings enhance
in the case $i\neq j$ for $pp \to \widetilde{\chi}_{i}^{0}\widetilde{\chi}_{j}^{0}$.
In addition, the maximum values of the \textit{K} factor are obtained in the region $\mu \lesssim 500$ GeV
and 400 $\lesssim M_2 \lesssim$ 1000 GeV for processes $pp \to \widetilde{\chi}_{1}^{0}\widetilde{\chi}_{1}^{0}$,
$\mu \lesssim 500$ GeV and $M_2 \lesssim$ 500 GeV for processes $pp \to \widetilde{\chi}_{1}^{0}\widetilde{\chi}_{2}^{0}$,
and  $\mu>M_{2}$ for process $pp \to \widetilde {\chi}_{2}^{0}\widetilde{\chi}_{2}^{0}$.
The LO and NLO cross sections for both $q \bar{q}$ annihilation and gg fusion are considerably determined by
the squark mass. When the squark mass increases by a factor of 4, the NLO cross section is pulled down by
around one and two orders of magnitude for $q \bar{q}$ annihilation and gg fusion, respectively.
However, the dependence of the LO cross section on the scale $\mu_0$ shows that it is
nearly independent of the scale $\mu_0$ for the above-mentioned reasons. We can also see that the \textit{K} factors decrease by about 11\% as the increment of the scale $\mu_0$ from 100 to 1000 GeV.

It should be underlined that there appear sizeable one-loop contributions to
the neutralino production, which considerably increase the
extracted bounds on the gaugino masses from the negative results for
these particles at the LHC. In our opinion these results will be helpful for investigations and analysis of the
different neutralino decay channels and for gaugino and Higgsino production
in the LHC and future hadron colliders.

\end{document}